\documentclass[final,seceqn]{elsarticle}
\usepackage{graphicx,lineno,hyperref,mathrsfs}
\usepackage{amsmath,amssymb,amsfonts,bm}
\usepackage{graphicx}
\usepackage[crop=off]{auto-pst-pdf}
\usepackage{epsfig}
\usepackage{pstricks}
\usepackage{amssymb}
\usepackage{color}
\usepackage{psfrag}
\definecolor{dblue}{rgb}{0.1,0.1,0.44}
\definecolor{dgreen}{rgb}{0.2 ,0.54, 0.2}
\journal{Nuclear Physics B}

 \renewcommand{\Im}{\mathrm{Im}}

\modulolinenumbers[10]

\begin{document}

\begin{frontmatter}

\title{Radiative corrections to the quark masses in the ferromagnetic Ising and Potts field theories}

\author{Sergei B. Rutkevich}
\ead{sergei.rutkevich@uni-due.de}
\address{Fakult\"at f\"ur Physik, Universit\"at Duisburg-Essen,  D-47058 Duisburg,Germany}
\begin{abstract}
We consider the Ising Field Theory (IFT), and the $3$-state  Potts Field Theory (PFT), which describe the scaling limits
of the two-dimensional lattice $q$-state Potts model with $q=2$, and $q=3$, respectively. At zero magnetic field $h=0$, 
both field theories are integrable away from the critical point, have $q$ degenerate vacua in the ferromagnetic
phase, and $q(q-1)$  particles of the same mass  -  the kinks interpolating between two different vacua. Application of 
a weak magnetic field induces confinement of kinks into bound states - the "mesons" (for $q=2,3$) consisting 
predominantly of  two kinks, and "baryons" (for $q=3$), which are essentially the  three-kink excitations. 
The kinks in the confinement regime are also called "the quarks". 
We review and refine
the Form Factor Perturbation Theory (FFPT), adapting it  to the analysis of the confinement problem in the limit of  small $h$,
and apply it to calculate  the corrections to the kink (quark) masses induced by the 
multi-kink fluctuations caused by the weak magnetic field. It is shown that the subleading 
third-order $\sim h^3$ correction to 
the kink mass  vanishes in the IFT. The leading second order $\sim h^2$ correction to the kink mass in the 3-state PFT is estimated by 
truncation  the infinite form factor expansion at the first term representing contribution of the two-kink fluctuations 
into the kink self energy. 
\end{abstract}


\end{frontmatter}

\linenumbers

\section{Introduction \label{Intr}}
Integrable models of statistical
mechanics and field theory \cite{Bax,Mussardo10} provide us with a very important source of information about the critical
behavior of condensed matter systems. Any progress in analytical solutions of such models is highly desirable, since it does not only yield exact information about the model itself but also about the whole universality class it represents. 
On the other hand, integrable models can
serve as zeroth-order  approximations in the perturbative analysis of their non-integrable deformations, providing a useful insight into a rich set of physical phenomena that never occur in integrable models: confinement of topological excitations, particle decay and inelastic scattering, false-vacuum decay, etc.

The Ising Field Theory (IFT) is the Euclidean quantum field theory that describes the scaling limit of the 
two-dimensional lattice Ising model near its phase transition point. 
Upon making a Wick rotation, the IFT can  be also viewed as a Lorentz-covariant field theory describing the dynamics of a one-dimensional quantum ferromagnet at zero temperature near its quantum phase
transition point \cite{Sach99}. The IFT is integrable at all temperatures for zero magnetic field $h=0$. 
Directly at the 
critical point  $T=T_c$, $h=0$ it reduces \cite{BPZ84} to the minimal conformal field theory $\mathcal{M}_3$,
which describes free massless Majorana fermions. These fermions acquire a nonzero mass $m\sim |T-T_c|$ at non-critical temperatures, but remain free at $h=0$. In the ordered phase $T>T_c$, the fermions are ordinary particles, while  in the ferromagnetic phase $T<T_c$ they 
become topological excitations - the kinks interpolating between two degenerate ferromagnetic vacua. Application of the magnetic field $h>0$ induces interactions between fermions and breaks the
integrability of the IFT at $T \neq T_c$. In the ordered phase $T<T_c$, it explicitly 
breaks also the degeneracy between ferromagnetic vacua. This  induces an  attractive long-range linear potential between the kinks, which  leads to their confinement into two-kink bound states. Due to the analogy with quantum chromodynamics, such bound states are often called "mesons", while  the kink topological 
excitations in such a
confinement regime are also called "quarks". In what follows, we shall synonymously use the terms ``kinks'' and ``quarks''. 

This mechanism of confinement  known as the McCoy - Wu scenario was 
first described for  the IFT  by these authors  \cite{McCoy78} in 1978, and 
attracted much interest in the last two decades. Recently it was experimentally observed and studied in 
one-dimensional quantum ferro- and anti-ferromagnets \cite{Coldea10,Mor14,Gr15,Wang15,Bera17}.
Since the IFT is not integrable at $h>0$, $m>0$ , different approximate techniques have been used for the theoretical understanding 
of the kink confinement in this model, such as  analytical perturbative expansions \cite{FonZam2003, FZ06,Rut05,Rut09} in the {\it weak confinement regime}
near the integrable direction $h=0$, and  numerical methods \cite{FZ06,LT2015}.  

The idea to use the  magnetic field as a perturbative  parameter characterizing a small deformation of an 
integrable massive  field theory was first realized  in the Form Factor Perturbation Theory (FFPT) introduced by  Delfino,  Mussardo, and Simonetti  
 \cite{Del96}. It turns out, however, that their original FFPT
cannot be applied directly to the kink confinement problem and requires considerable modification. 
The reason
is that even an arbitrarily weak long-ranged confining interaction leads to qualitative changes of the particle
content at the confinement-deconfinement transition: isolated kinks cannot exist any more  in the presence of the magnetic field, and 
the mass spectrum $M_n(m,h)$, $n=1,2,\ldots$ of their bound  states (the mesons), become dense in the interval $2m<M_n<\infty$
in the limit $h\to+0$.
This in turn makes straightforward perturbation theory
based on the adiabatic hypothesis unsuitable. { A different, non-perturbative technique to study 
the IFT meson mass spectrum was developed by Fonseca and Zamolodchikov \cite{FonZam2003}. This technique
is based on the Bethe-Salpeter equation, which was derived for the IFT 
in \cite{FonZam2003}   in the two-quark approximation.}
The latter approximation 
implies that  at small magnetic fields $h\to+0$,   the meson eigenstate 
\begin{equation}\label{mesonwf}
|\Psi_P\rangle=|\Psi_P^{(2)}\rangle+|\Psi_P^{(4)}\rangle+|\Psi_P^{(6)}\rangle+\ldots
\end{equation}
 of the IFT Hamiltonian, with $P$ being the meson momentum, is approximated by the two-quark component
\begin{equation}
|\Psi_P^{(2)}\rangle=\frac{1}{2}\int_{-\infty}^\infty\frac{dp_1}{2\pi}\frac{dp_2}{2\pi}\delta(p_1+p_2-P)
\,\Psi_P(p_1,p_2)\,|p_1p_2\rangle,
\end{equation}
neglecting the multi-quark contributions represented by further terms in the right-hand side of \eqref{mesonwf}. Here $p_{1}, p_2$ denote the momenta of two quarks coupled into a meson.

{It was shown in  \cite{Rut09},
that the FFPT can be  modified to adapt it to the confinement problem, if one takes into account the long-range attractive potential already at zeroth order and applies a certain $h$-dependent unitary transform 
in the Fock space of the  free IFT.  Such  a modified FFPT incorporates the Bethe-Salpeter 
equation in its leading order. This perturbative technique  can be effectively used in 
the weak confinement regime  $h\to+0$ despite  the 
break of the adiabatic hypothesis at  the confinement-deconfinement transition at $h=0$.} 

Two kinds of asymptotic expansions for the meson masses $M_n(m,h)$ have been obtained for 
the IFT in the weak confinement regime $h\to+0$. The {\it low energy expansion} 
\cite{McCoy78,FonZam2003, FZ06,Rut09}
in fractional powers of 
$h$ describes the initial part of the meson mass spectrum, while the {\it semiclassical expansion} 
\cite{FZ06,Rut05, Rut09} in integer powers of $h$ describes the meson masses $M_n(m,h)$ with $n\gg1$.
High accuracy of both expansions has been established \cite{FZ06,LT2015} by comparison with the IFT meson mass spectra calculated by direct numerical methods based on the Truncated Conformal Spaced Approach \cite{YuZam90, YuZam91}. 

The leading terms in the  low energy and semiclassical expansions  can be gained from the Bethe-Salpeter equation. This indicates \cite{FZ06}, that the two-quark approximation is asymptotically exact to the leading order in $h\to 0$. It was shown \cite{FonZam2003, FZ06}, 
however, that starting  from the second order in $h$ in both 
low energy and semiclassical expansions, one must take into account the mixture of four-quark, six-quark, etc.  configurations in the meson state \eqref{mesonwf}. 

The leading multi-quark correction to the meson masses in the IFT was obtained by Fonseca and Zamolodchikov \cite{FZ06}. This correction is of order $h^2$, and originates from the renormalization of the quark mass. The third-order 
$\sim h^3$  multi-quark corrections  to the IFT meson masses have so far only partly been known. 
These corrections arise from contributions of three effects.
\begin{itemize}
\item Renormalization of the long-range attractive force between the neighboring kinks (the 'string-tension')
of order $h^3$  which was determined in  \cite{FZ06}. 
\item
Multi-quark fluctuations modify the regular part of the Bethe-Salpeter kernel, which is
responsible for the pair interaction between quarks at short distances. The corresponding contribution $\sim h^3$ to 
the meson masses was found in \cite{Rut09}.
\item The  radiative corrections of the quark mass of the third-order in $h$, which was  unknown.
\end{itemize}

The first aim of this paper is to complete the calculation of the 
meson mass spectrum in the IFT in the weak confinement regime $h\to+0$ to third order in $h$. 
To this end, we  review and { further}  modify the form factor perturbative technique  developed for the confinement 
problem in  \cite{Rut09}. The FFPT contains a well known problem  caused  by the so-called kinematic singularities in the matrix elements of the spin operator. Merging of  such singularities in the integrals arising in the FFPT leads to ill-defined quantities like $\delta(0)$, or $\delta(p)/p$. We propose a  
consistent regularization procedure that allows one to  perform  high-order FFPT calculations in a controlled fashion 
 avoiding  ill-defined quantities in 
intermediate expressions. 
 The key idea is to replace the uniform magnetic field in the Hamiltonian of the 
infinite system 
by its nonuniform 
counterpart switched on in a { finite} interval of the length $R$, to perform all calculations at a 
large but finite $R$, and to proceed to the limit $R\to\infty$ afterwards. To verify the efficiency of this regularization procedure, we  use
 it to reproduce several well-known results and 
to obtain some new ones  for the scaling limit of the Ising model.  Then we apply the same procedure
to calculate the third-order radiative correction to the quark mass in the ferromagnetic IFT
showing that it vanishes.

The mechanism of confinement outlined above is quite common in two-dimensional quantum field theories,
that are
invariant under some discrete symmetry group and display a continuous order-disorder phase transition. 
If such a model has several degenerate vacua in the ordered phase, the application of an external
field  typically leads to confinement of  
kinks interpolating between  different vacua. {  Realizations of this scenario in different two-dimensional
models have been the subject of considerable interest in recant years \cite{DelMus98,Del08,Mus07,
Mus08,MusTak09}. In this paper we shall address to some aspects of the confinement problem in the  three-state Potts Field Theory (PFT).}
 
The three-state PFT represents the scaling limit of the two-dimensional lattice three-state Potts model \cite{Bax,Wu82}. 
At zero magnetic field, it is invariant under the permutation group ${\mathbb S}_3$  and displays 
the continuous order-disorder phase transition. 
It was shown by Dotsenko  \cite{DOTSENKO198454}, that the conformal field theory corresponding 
to the critical point of the three-state Potts model can be identified as the minimal unitary model  $\mathcal{M}_5$.  In the ordered phase at zero magnetic field, the three-state PFT
has three degenerate vacua and six kinds of massive particles of the same 
mass - the kinks ('quarks') $K_{\mu \nu}$   interpolating between  vacua $|0\rangle_\mu$ and $|0\rangle_\nu$, where  $\mu,\nu\in \mathbb{Z} \,{{\rm mod}\,3}$.  The three-state PFT is integrable at 
zero magnetic field  \cite{CZ92}, and the quark scattering matrix is exactly known \cite{KOBERLE1979209}.
This scattering matrix is non-trivial, which indicates that the quarks in the three-state PFT are not free at zero magnetic field,  but strongly  interact with each other at small distances, in contrast to the IFT. 
The form factors of the physically relevant operators in the massive three-state PFT were determined by 
Kirillov and Smirnov \cite{KS88}.

Application of the magnetic field $h\ne0$ breaks integrability of the PFT and leads to confinement of quarks. 
The quark bound states  in the $q$-state PFT in the confinement regime were classified by Delfino and 
Grinza \cite{Del08}, who also showed that besides the mesonic (two-quark) bound states, the baryonic 
(three-quark) bound states are allowed at $q=3$.  First numerical calculations of the 
meson and baryon mass spectra in the $q$-state PFT were described in 
\cite{Del08,LTD}.  The meson masses in the $q$-state PFT in the
weak confinement regime were analytically calculated  to  leading order in $h$ in  \cite{RutP09}, where the generalization of the IFT 
Bethe-Salpeter equation to the PFT was also
described. The masses of several lightest baryons in the three-state PFT in the leading order in $h$ have been calculated in \cite{Rut15B}. Analytical predictions of \cite{RutP09,Rut15B} 
for the meson and baryon masses in the  three-state PFT were confirmed in direct numerical 
calculations performed by  Lencs{\'e}s and   Tak{\'a}cs  \cite{LT2015}. 

The second subject of the present paper is to estimate  the second-order radiative  correction 
to the quark masses in the 3-state PFT in the weak confinement regime. 
This correction to the quark mass gives rise to the multi-quark corrections to the 
meson and baryon masses in  second order in $h$. 
Starting from 
the  Lehmann expansion for the quark mass radiative  correction,
 we calculate its first term representing the quark self-energy diagram with two virtual quarks in the
 intermediate state. 

The remainder of this paper is organized as follows.
In the next section we start with recalling some well-known properties of the $q$-state Potts model on the square lattice, and then describe briefly its scaling limit in the case $q=3$, and zero magnetic field. 
In Section \ref{Is} we review the FFPT ad{a}pted in \cite{Rut09} 
to the confinement problem in the IFT. 
We further improve this FFPT technique in order to regularize the products of singular matrix elements of the spin operator which arise in this method. 
 We then apply the improved version of the FFPT  to recover some well-known results and to obtain several new ones   for the IFT. In Section \ref{FF} we describe the form factors of the disorder spin operators 
in the  three-state PFT at zero magnetic field in the paramagnetic phase, which were found by Kirillov  and Smirnov \cite{KS88}. 
Applying the duality transform to these form factors, we obtain the matrix elements of the order spin operators in the ferromagnetic three-state PFT between the one- and two-quark states. These matrix elements are used in Section \ref{SOPFT} to estimate the second-order correction to 
the quark mass in the latter model in the presence of a weak magnetic field. Concluding remarks are
given in Section \ref{Conc}. Finally, there are four
appendixes describing technical details of some of the required
calculations.
\section{Potts Field Theory \label{PFTsec}}
In this section we following \cite{Del08}   review  
some well known properties of the $q$-state Potts model on the square lattice, and then proceed to its scaling limit. 

Consider the two-dimensional  square lattice ${\mathbb Z}^2$ 
and associate with each lattice site $x\in {\mathbb Z}^2 $ a discrete spin variable 
$s(x) =1,2,\ldots,q$. The model
Hamiltonian is  defined as
\begin{equation}
{\cal E}=-\frac{1}{T} \sum_{<x,\,y>} \delta_{s(x),s(y)}-H\sum_x
  \delta_{s(x),q}.
\label{HamP}
\end{equation}
Here the first summation is over nearest neighbour pairs, $T$ is the 
temperature, $H$ is the external magnetic field
applied along the $q$-th direction, and $\delta_{\alpha,\alpha'}$ is the Kronecker symbol.
 At $H=0$, the Hamiltonian (\ref{Ham}) is invariant under the permutation group ${\mathbb S}_{q}$; at $H\ne0$  
the symmetry group reduces to ${\mathbb S}_{q-1}$. At $q=2$, model \eqref{HamP} reduces to the Ising model.

The order parameters $\langle\sigma_\alpha\rangle$ can be associated with the variables
\[
\sigma_\alpha(x) =\delta_{s(x),\alpha}-\frac{1}{q}, \quad \alpha=1,\ldots,q. 
\]
The parameters $\langle\sigma_\alpha\rangle$ are not independent, since
\begin{equation}
\sum_{\alpha=1}^q\sigma_\alpha(x)=0.
\end{equation}

Two complex spin variables $\sigma(x)$ and $\bar{\sigma}(x)$ 
defined by the relations
\begin{align}
\sigma(x)=\exp [2 \pi i s(x)/q]=\sum_{\alpha=1}^q \exp (2 \pi i \alpha/q)\,\sigma_\alpha(x),\\
\bar{\sigma}(x)=\exp [-2 \pi i s(x)/q]=\sum_{\alpha=1}^q \exp (-2 \pi i \alpha/q)\,\sigma_\alpha(x),
\end{align} 
are useful in proceeding to the continuous limit.

At  zero magnetic field, the  model undergoes a ferromagnetic phase transition at the critical temperature
\begin{equation}
  T_c=\frac{1}{ \log(1+\sqrt{q})}.
\end{equation}
This transition  is continuous for $2\le q\le 4$.  
The ferromagnetic low-temperature phase at zero field 
is $q$-times degenerated. { 
The  Potts model \eqref{HamP}   at $H=0$ possesses the dual symmetry, which generalizes 
 the Kramers-Wannier duality of the Ising model. This symmetry connects the properties  of the model in the
ordered and disordered phases. By duality, the partition functions of the zero-field Potts model  coincide 
at the temperatures $T$ and $\tilde{T}$, provided
\[
\left(e^{1/T}-1 \right)\left(e^{1/\tilde{T}}-1 \right)=q.
\]
}
For a review of many other known properties of the Potts model see \cite{Wu82,Bax}.

The scaling limit of 
the model (\ref{HamP}) at $H\to 0$, $T\to T_c$, and $q\in [2,4]$  is described by the 
Euclidean action \cite{Del08} 
\begin{equation} \label{AP}
{\cal A}^{(q)}= {\cal A}_{CFT}^{(q)} -\tau\int d^2x\,{\mathfrak e}(x)-
h\int d^2x\,\sigma_q(x)\,\,,
\end{equation}
Here $x$ denotes the points of the plane $\mathbb{R}^2$ having the cartesian coordinates 
$(\rm{x},\rm{y})$.
The first term ${\cal A}_{CFT}^{(q)}$ corresponds to the conformal field theory, 
which is associated with the critical point. 
Its central charge $c(q)$ takes the value 
\begin{equation}
c(q)=1-\frac{6}{t(t+1)}, \quad {\rm where}\;\;\sqrt{q}=2\sin \frac{\pi(t-1)}{2(t+1)}.
\end{equation}

The fields ${\mathfrak e}(x)$ (energy density)
and $\sigma_q(x)$ (spin density) are characterized by the 
scaling dimensions 
\[ 
X_{\mathfrak e}^{(q)}=\frac{1}{2}\left(1+\frac{3}{t}\right),  \quad \quad 
X_\sigma^{(q)}=\frac{(t-1)(t+3)}{8 t(t+1)}. 
\]
The parameters $\tau\sim (T-T_c)$ and $h\sim H$ are proportional
to the deviations of the  temperature and the magnetic field
from their critical point values. At $h=0$ and $\tau\ne 0$ the field theory \eqref{AP} is integrable, 
i.e. it has infinite number of integrals of motion and a
factorizable scattering matrix \cite{CZ92}.

In the rest of this section we shall  concentrate on the  $q=3$ Potts field theory.  
The simpler and better studied   Ising case  corresponding to $q=2$ will be
discussed in Section \ref{Is}.

\subsection{Disordered phase at $h=0$}
The model has a unique ground state $|0\rangle_{  par}$ in the disordered phase, at $\tau>0$ and $h=0$.   
The particle content of the model consists of a massive scalar particle and its antiparticle.
Their  momentum $p$ and energy  
\begin{equation}\label{omff}
\omega(p)=\sqrt{p^2+m^2}
\end{equation}
can be 
conveniently parametrized by the rapidity $\beta$, 
\begin{equation}\label{disp}
p(\beta)=m \sinh\beta, \quad \omega(\beta)=m \cosh\beta. 
\end{equation}
  Here  $m\sim  {\tau}^{5/6}$ is the particle mass.  
 
The  space of states is generated by the Faddeev-Zamolodchikov creation/annihilation operators $Z_\varepsilon^*(\beta)$, 
$Z_\varepsilon(\beta)$, where the index $\varepsilon=\pm1$ distinguishes  particles ($\varepsilon=1$) and 
antiparticles ($\varepsilon=-1$). These operators satisfy the following equations
\begin{align}\label{ZZcom}
Z_{\varepsilon_1}(\beta_1)\,Z_{\varepsilon_2}(\beta_2)=S_{\varepsilon_1,\varepsilon_2}(\beta_1-\beta_2)Z_{\varepsilon_2}(\beta_2)\,Z_{\varepsilon_1}(\beta_1),\\ {\label{ZZ2}
Z_{\varepsilon_1}^*(\beta_1)\,Z_{\varepsilon_2}^*(\beta_2)=S_{\varepsilon_1,\varepsilon_2}(\beta_1-\beta_2)Z_{\varepsilon_2}^*(\beta_2)\,Z_{\varepsilon_1}^*(\beta_1) },\\
Z_{\varepsilon_1}(\beta_1)\,Z_{\varepsilon_2}^*(\beta_2)=S_{\varepsilon_2,\varepsilon_1}(\beta_2-\beta_1)Z_{\varepsilon_2}^*(\beta_2)\,Z_{\varepsilon_1}(\beta_1)+\delta_{\varepsilon_1\varepsilon_2}
\delta(\beta_1-\beta_2),\label{FZ2}
\end{align}
where
\begin{eqnarray}\label{S}
&&S_{-1,-1}(\beta)=S_{1,1}(\beta)=\frac{\sinh[(\beta+2\pi i/3)/2]}{\sinh[(\beta-2\pi i/3)/2]},\\
&&S_{1,-1}(\beta)=S_{-1,1}(\beta)=S_{1,1}(i\pi-\beta). \nonumber
\end{eqnarray}
Equation \eqref{FZ2} implies that  the one-particle states are normalized as
\begin{equation}\label{norm3P}
\phantom{x}_{ par} \langle 0|Z_{\varepsilon_1}(\beta_1)Z_{\varepsilon_2}^*(\beta_2)|0\rangle_{ par} =
\delta_{\varepsilon_1\varepsilon_2}
\delta(\beta_1-\beta_2).
\end{equation}
The two-particle scattering amplitudes \eqref{S} were found by K\"oberle and Swieca \cite{KOBERLE1979209}.
The generators of the 
permutation group $\mathbb{S}_3\approx \mathbb{Z}_3\times \mathbb{Z}_2$ act 
on the paramagnetic vacuum  and particles as follows
\begin{align}
\Omega |0\rangle_{ par}  =|0\rangle_{ par} ,\quad C |0\rangle_{ par} =|0\rangle_{ par} , \\
\Omega Z_{\varepsilon}^*(\beta)\Omega^{-1}= \upsilon^{\varepsilon} Z_{\varepsilon}^*(\beta), \\
C Z_{\epsilon}^*(\beta) C^{-1}=Z_{-\epsilon}^*(\beta).
\end{align}
Here $\upsilon=\exp(2\pi i/3)$, $\Omega$ is the generator of the
cyclic permutation group $\mathbb{Z}_3$, $\Omega^3 =1$,
 $C$ is the charge conjugation, $C^2=1$.

{ The vector  space $\mathcal{L}_{par}$ of paramagnetic states 
is spanned by the paramagnetic vacuum $|0\rangle$, 
and the $n$-particle vectors 
\begin{equation}\label{rapibas}
|\beta_n,\ldots,\beta_2,\beta_1\rangle_{\varepsilon_n,\ldots,\varepsilon_2,\varepsilon_1} \equiv 
Z_{\varepsilon_n}^*(\beta_n)\ldots Z_{\varepsilon_2}^*(\beta_2) Z_{\varepsilon_1}^*(\beta_1)  |0\rangle_{par},
\end{equation}
with $n=1,2,\ldots.$ Corresponding to \eqref{rapibas} bra-vector is denoted as
\[
\phantom{x}_{\varepsilon_1,\varepsilon_2,\ldots,\varepsilon_n}\langle\beta_1,\beta_2,\ldots, \beta_n|\equiv 
\phantom{x}_{ par} \langle 0|Z_{\varepsilon_1}(\beta_1) Z_{\varepsilon_2}(\beta_2) \ldots  Z_{\varepsilon_n}(\beta_n)  .
\]

Let us denote by $\mathcal{L}_{sym}$ the subspace of $\mathcal{L}_{par}$ spanned by the vacuum $|0\rangle$
and vectors \eqref{rapibas}, for which $\sum_{j=1}^n\epsilon_j =0\,{\rm mod}\, 3$.
Operator $\Omega$ acts as the identity operator on the subspace $\mathcal{L}_{sym}$.

The $n$-particle vectors \eqref{rapibas} are not linearly independent, 
but satisfy a number of linear relations, which are imposed 
on them by  the commutation relations \eqref{ZZ2}. For example, 
\begin{equation}\label{examp}
|\beta_1,\beta_2\rangle_{\varepsilon_1,\varepsilon_2}=S_{\varepsilon_1,\varepsilon_2}(\beta_1-\beta_2)
|\beta_2,\beta_1\rangle_{\varepsilon_2,\varepsilon_1}.
\end{equation}
The "in"-basis in the $n$-particle subspace  
$\mathcal{L}_{par}^{(n)}$  of  $\mathcal{L}_{par}$
 is formed by the vectors of the form \eqref{rapibas} with $\beta_n>\beta_{n-1}>\ldots>\beta_1$,
and the  "out"-basis in the same subspace $\mathcal{L}_{par}^{(n)}$ is formed by the vectors 
\eqref{rapibas} with $\beta_n<\beta_{n-1}<\ldots<\beta_1$.

}

Reconstruction of the matrix elements of local operators between such basis states in integrable models
is the main subject of the form factor bootstrap program  
\cite{smirnov1992form}. For the three-state PFT, this program was realized by Kirillov and Smirnov in  \cite{KS88}, where the explicit representations for the form factors of the main 
operators naturally arising in this model were obtained. We postpone the discussion of these results to 
Section \ref{FF}.
\subsection{Ordered phase at $h=0$ \label{FERR}}
In the low temperature phase $\tau<0$, the ground state $|0\rangle_\mu$, $\mu={0},1,2\,
 {\rm mod}\, 3$ is three-fold degenerate at $h=0$.  The elementary excitations are topologically charged being represented by  six kinks 
{ $|K_{\mu\nu}(\beta)\rangle$},  $\mu,\nu\in \mathbb{Z}\,{{\rm mod}\, 3}$ interpolating between two different vacua $|0\rangle_\mu$ and $|0\rangle_\nu$.  
These kinks are massive relativistic particles with the 
mass $m\sim\,(-\tau)^{5/6}$.

{
The  generators of the symmetry group $\mathbb{S}_3$ act on the vacua and  one-kink states as follows,
\begin{eqnarray}
&&{\tilde{\Omega}} |0\rangle_\mu=|0\rangle_{\mu+1}, \\
&&{\tilde{C}} |0\rangle_\mu=|0\rangle_{-\mu},\\
&&{\tilde{\Omega}} |K_{\mu\nu}(\beta)\rangle=|K_{\mu+1,\nu+1}(\beta)\rangle,\label{OmK}\\
&&{\tilde{C}} |K_{\mu\nu}(\beta)\rangle=|K_{-\mu,-\nu}(\beta)\rangle.
\end{eqnarray}

The subspace  $\mathcal{L}_{fer}^{(n)}$ of the $n$-kink states in the ferromagnetic space  $\mathcal{L}_{fer}$ 
is spanned by the vectors 
\begin{equation}\label{kinkS}
|K_{\mu_n \mu_{n-1}}(\beta_n)\ldots K_{\mu_2 \mu_{1}}(\beta_2) K_{\mu_1 \mu_{0}}(\beta_1)\rangle.
\end{equation}
Corresponding bra-vector is denoted as
\[
\langle K_{\mu_0 \mu_{1}}(\beta_1) K_{\mu_1 \mu_{2}}(\beta_2) \ldots 
K_{\mu_{n-1} \mu_{n}}(\beta_1)|.
\]
The $n$-kink  states \eqref{kinkS} are called topologically neutral, 
if $\mu_n=\mu_0 $, and topologically charged otherwise.
We  denote by $\mathcal{L}_0$ the topologically neutral subspace of $\mathcal{L}_{fer}$ spanned by the 
ferromagnetic vacuum $|0\rangle_0$, and vectors \eqref{kinkS} with $\mu_n=\mu_0=0 $.

The Kramers-Wannier duality of the square-lattice Potts model \cite{Bax,Wu82} manifests itself also in the quantum 
Potts spin chain model \cite{Tak13}, and in the scaling PFT  at  and beyond the critical  point \cite{DOTSENKO198454,CZ92} . 
Roughly speaking, the duality symmetry in the latter case can be viewed as  the kink-particles correspondence \cite{Del08,Tak13}
\begin{eqnarray*}
&&|K_{10}(\beta) \rangle, \;|K_{21}(\beta) \rangle, \;|K_{02}(\beta) \rangle   \quad \longleftrightarrow  \quad|\beta \rangle_1,\\
&&|K_{01}(\beta) \rangle,  \;|K_{12}(\beta) \rangle, \;|K_{20}(\beta) \rangle   \quad \longleftrightarrow \quad |\beta\rangle_{-1}
\end{eqnarray*} 
between the elementary excitations in the ferromagnetic and paramagnetic phases. 
To be more precise, let us define the duality  transform $\mathcal{D}$  as a linear mapping 
$\mathcal{L}_0\to \mathcal{L}_{sym} $
determined
   by the following relations
\begin{eqnarray}
&&\mathcal{D}\, |0\rangle_0=|0\rangle_{ par},\\\label{dualV}
&&\mathcal{D} |K_{\mu_{n},\mu_{n-1}}(\beta_n),
\ldots, K_{\mu_{1},\mu_0}(\beta_1)\rangle=|\beta_n,\ldots,\beta_1\rangle_{\epsilon_n,\ldots,\epsilon_1},
\end{eqnarray}
where
\begin{equation}
\epsilon_j=\begin{cases}
1, \;\; {\rm if} \;\;\mu_j-\mu_{j-1}=1\,{\rm mod\,} 3,\\
-1, \; {\rm if} \;\;\mu_j-\mu_{j-1}=-1\,{\rm mod\,} 3,
\end{cases}
\end{equation}
and  $\mu_n =\mu_0 =0 $.

The Kramers-Wannier duality of the PFT 
requires that the mapping  $\mathcal{D}$ must be  unitary, i.e. 
the inverse mapping $\{\mathcal{D}^{-1}|\mathcal{D}^{-1}:\mathcal{L}_{sym}\to \mathcal{L}_{0} \} $ must
exist, and $\mathcal{D}^{-1}=\mathcal{D}^{\dagger}$. These requirements lead to a number of linear 
relations between the $n$-kink states \eqref{kinkS}. For example,  acting 
on the equality 
\[
|\beta_1,\beta_2\rangle_{1,-1}=S_{1,-1}(\beta_1-\beta_2)|\beta_2,\beta_1\rangle_{-1,1}
\]
[following from \eqref{examp}] by the mapping $\mathcal{D}^{-1}$, one obtains, 
\[
|K_{02}(\beta_1) K_{20}(\beta_2) \rangle=S_{1,-1}(\beta_1-\beta_2)|K_{01}(\beta_2) K_{10}(\beta_1) \rangle.
\]
Application of the same procedure to the $n$-particle states \eqref{kinkS} leads to the Faddeev-Zamolodchikov 
commutation relations
\begin{subequations}\label{KK}
\begin{align}
K_{\mu\nu}(\beta_1)K_{\nu\gamma}(\beta_2)= S_{1,1}(\beta_1-\beta_2)K_{\mu\nu}(\beta_2)K_{\nu\gamma}(\beta_1),\\
K_{\mu\nu}(\beta_1)K_{\nu\mu}(\beta_2)= S_{1,-1}(\beta_1-\beta_2)K_{\mu\rho}(\beta_2)K_{\rho\mu}(\beta_1),
\end{align}
\end{subequations}
where $\rho\ne\nu$.
According to the conventional agreement \cite{ZZ79}, notations  $K_{\alpha\alpha'}(\beta_j)$  in the 
above relations can be understood as the formal 
non-commutative  symbols  representing the kinks in the $n$-kink states
 \eqref{kinkS}. 

Relations \eqref{KK} describe the two-kink scattering processes in the ferromagnetic phase. 
Due to the PFT  dual symmetry, they are characterized  by the same scattering amplitudes, as the two-particle
scattering in the paramagnetic phase.  Furthermore, the scattering theories in the high- and low-temperature phases are equivalent. 
Such duality arguments
}
 can be also extended to the matrix elements of   physical operators. In particular, the matrix elements of the order spin operators in the ferromagnetic phase can be expressed in terms of the form factors of the disorder spin operators \cite{FRADKIN19801}
in the paramagnetic phase. { We shall return to this issue in Section \ref{FF}.}

\section{Quark mass in the ferromagnetic IFT \label{Is}}
The IFT action $A_{IFT}\equiv\mathcal{A}^{(2)}$ is defined  by equation \eqref{AP} with $q=2$.  
The conformal field theory $\mathcal{A}_{CFT}^{(2)}$ associated with the critical point 
is the minimal model $\mathcal{M}_3$, which  contains
free massless Majorana fermions \cite{BPZ84}. These fermions acquire a mass $m\sim |\tau|$, as the temperature deviates from the critical point. They remain free at $h=0$. However,  application of a magnetic field $h>0$ induces  interaction between the fermions. 
The  Hamiltonian corresponding  to the action $A_{IFT}$ can be written as \cite{Rut09}
\begin{eqnarray}
&&\mathcal{H} = \mathcal{H}_0 +h\,V, \label{Ham}\\
\textrm{where   }\;\;\;&&\mathcal{H}_0=\int_{-\infty}^\infty \frac{d p}{2 \pi} 
\,\omega(p)\, {\bf a}^\dagger (p) \, {\bf a}(p),  \label{H0}\\
&&V=-\int_{-\infty}^\infty  d\rm{x}\,\sigma(\rm{x}),   \label{V}
\end{eqnarray}
and  $\omega(p)$ is
the spectrum   \eqref{omff} of free  fermions. These fermions 
are ordinary spinless particles in the disordered phase $\tau>0$, and topologically-charged kinks
interpolating between two degenerate vacua in the ordered phase $\tau<0$.
Fermionic operators $ {\bf a}^\dagger (p') , \,{\bf a}(p)$  obey the canonical 
anticommutational relations
\[
\{ {\bf a}(p) , {\bf a}^\dagger (p') \} =2 \pi \,\delta(p-p'), \quad 
 \{ {\bf a}(p), {\bf a} (p') \} = \{ {\bf a}^\dagger (p) ,{\bf a}^\dagger (p') \}= 0.  
\]
Commonly used are also fermionic operators $a(\beta),\, a^\dagger(\beta)$, 
corresponding to the rapidity variable $\beta={\rm arcsinh}(p/m)$:
\begin{equation}\label{abeta}
a(\beta)= \omega(p)^{1/2}\, {\bf a}(p), \;a^\dagger(\beta)=\omega(p)^{1/2}\, {\bf a}^\dagger (p). 
\end{equation}
The notations 
\begin{eqnarray*}
|p_1, \dots ,p_N \rangle = {\bf a}^\dagger (p_1) \dots  {\bf a}^\dagger (p_N) |0\rangle, \quad\;\;
\langle p_1,\ldots,p_N|=\langle 0|{\bf  a}(p_1) \dots  {\bf  a}(p_N) , \\
| \beta_1,\dots, \beta_N \rangle = a^\dagger(\beta_1) \dots a^\dagger(\beta_N)|0\rangle, \quad\quad
\langle \beta_1,\dots ,\beta_N|=\langle 0|  a(\beta_1) \dots a(\beta_N)
\end{eqnarray*}
for the fermionic basis states with definite momenta will be used. 

The order spin operator $ \sigma({\rm x}) =\sigma( {\rm x} ,{\rm y})|_{{\rm y}=0}$ in 
the ordered phase $\tau<0$  
is  completely characterized by the matrix elements
$ \langle \beta_1,\ldots,\beta_K|\sigma(0) |\beta'_1,\ldots,\beta'_{N}\rangle $, whose
explicit expressions are well known  
\cite{Berg79,FonZam2003}, see  equation  (2.14) in \cite{FonZam2003}. These matrix elements are different from zero only if  $K+N=0\,({\rm mod}\,2)$.
The matrix elements with $K+N=2$  read as
\begin{align}\label{fIs}
\langle p |  \sigma({\rm x})|k\rangle = \frac{  i\,\bar{\sigma}\, \exp[ i  {\rm x}(k-p)]   }{p-k}\,
\frac{ \omega(p)+\omega(k) }{[\omega(p)\omega(k)]^{1/2}},\\
\langle 0 | \sigma({\rm x}) |k_1 k_2\rangle = \frac{  i\,\bar{\sigma}\, \exp[ i  {\rm x}(k_1+k_2)]   }{k_1+k_2}\,
\frac{ \omega(k_1)-\omega(k_2) }{[\omega(k_1)\omega(k_2)]^{1/2}},\label{fIs2}\\
\langle k_1 k_2 | \sigma({\rm x}) |0\rangle = \frac{  i\,\bar{\sigma}\, \exp[- i  {\rm x}(k_1+k_2)]   }{k_1+k_2}\,
\frac{ \omega(k_1)-\omega(k_2) }{[\omega(k_1)\omega(k_2)]^{1/2}},\label{fIs3}
\end{align}
where $\bar{\sigma}=\bar{s} |m|^{1/8}$  is the zero-field vacuum 
expectation value of the order field (spontaneous magnetization), and 
\begin{equation}
\bar{s}=2^{1/12}e^{-1/8}A^{3/2}=1.35783834...,
\end{equation}
where $A=1.28243...$ stands for the  Glaisher's constant.  
The matrix elements of the order spin operator with $K+N>2$ can be determined from 
\eqref{fIs}-\eqref{fIs3} by means of the  Wick expansion. 
For real $p$ and $k$, the {"kinematic"} pole at $p=k$ in \eqref{fIs} is understood in the 
sense of the Cauchy principal value
\begin{equation}\label{Cau}
\frac{1}{p-k} \to\mathcal{P} \frac{1}{p-k} \equiv 
\frac{1}{2}\left(\frac{1}{p-k+i 0}+ \frac{1}{p-k-i 0}
\right).
\end{equation}

The field theory defined by the Hamiltonian \eqref{Ham}-\eqref{V} is not integrable for 
generic $m>0$ and $h>0$, but admits exact solutions along the lines $h=0$ and $m=0$. 
The line $h=0$ corresponds to the Onsager's solution \cite{Ons44}, whose scaling limit describes 
free massive fermions. Integrability of the IFT along the line $m=0$, $h\ne 0$ was established by Zamolodchikov \cite{ZamH}. 

Close to integrable directions, it is natural to treat the non-integrable quantum field
theories as  deformations of  integrable ones. As it was mentioned in the Introduction,
realization of this idea leads to the FFPT, whose original version  \cite{Del96}, however, 
cannot be applied directly to the confinement problem 
since the magnetic field changes the particle content of the theory at arbitrary small $h>0$.
The problem manifests itself already in the naive first-order correction formula for the kink mass  \cite{Del96}
\begin{equation}
\delta^{(1)} m= -
\lim_{p\to 0} \lim_{k\to p}h\, \langle p |\sigma(0)|k\rangle,
\end{equation}
which is infinite due to the kinematic pole in the matrix element \eqref{fIs} of the spin operator.  To avoid this problem, a modified version of the FFPT was developed in \cite{Rut09}.
Since it is substantially used in this section, it will be helpful  to recall here its main issues. 

The kea idea of the modified  FFPT is to absorb a part of the interaction into the unitary operator 
$U(h)$, for which the formal expansion  in powers of $h$ is postulated,
\begin{equation}\label{Uh}
U(h)=1+\sum_{n=1}^\infty h^n\,\mathcal{F}_n. 
\end{equation}
This operator  has been used to define creator and annihilator operators for the  "dressed" fermions, 
\begin{equation}\label{dr}
\underline{{\bf a}}(p)=U(h)^{-1} \,{{\bf a}}(p) \,U(h), \quad   
\underline{{\bf a}}^\dagger(p)=U(h)^{-1}\, {{\bf a}}^\dagger(p) \,U(h),
\end{equation}
which are underlined to distinguish them from the "bare" ones. Similarly, the dressing unitary transform
is defined for arbitrary operators and states, 
\begin{equation*}
\underline{A}=U(h)^{-1} \,A\,U(h) , \qquad  |\underline{\Phi}\rangle =U(h)^{-1} |{\Phi}\rangle.
\end{equation*}

It was required in \cite{Rut09} that the number of dressed fermions conserves in the evolution
defined by the Hamiltonian \eqref{Ham}-\eqref{V}, i.e.
\begin{equation}\label{NH}
[\underline{N},\mathcal{H}]=0,
\end{equation}
where
\[
\underline{N}=\int_{-\infty}^\infty \frac{d p}{2\pi}\,\underline{{\bf a}}^\dagger(p)\,\underline{{\bf a}}(p).
\]
It was required, further, that
operators $\mathcal{F}_n$ change the number of dressed fermions, i.e.
\begin{equation} \label{mcFd}
\langle \underline{p}|\mathcal{F}_n|\underline{k}\rangle =0 \quad \textrm{for} \quad  n(p)= n(k) .
\end{equation}
Here the shortcut notations  
$
| \underline {k}\rangle =|\underline {k_1,...,k_{n(k)}}\rangle,$
$\langle \underline {p}|=
\langle \underline {p_1,...p_{n(p)}}|$, 
have been used. 

Conditions \eqref{NH}, \eqref{mcFd} together with the unitarity requirement 
\begin{equation}\label{U}
U(h)U(h)^{-1}=1, 
\end{equation}
allow one 
to determine the coefficients $\mathcal{F}_n$ in the expansion \eqref{Uh}. In particular, the matrix elements of the first one 
read as
\begin{eqnarray} \label{mcF}
\langle \underline{p}|\mathcal{F}_1|\underline{k}\rangle =
\frac{\langle {p}|{V}|{k}\rangle}
{\omega(p)-\omega(k)}, \quad \textrm{for} \quad  n(p)\ne n(k) ,
\end{eqnarray}
where we again use the abbreviation  $\omega (q)\equiv \omega (q_1)+...+\omega(q_{n(q)})$.

Note that the matrix element \eqref{mcF} diverges at the hyper-surface determined by the 
"resonance relation"
\begin{equation} \label{RES}
\omega(p_1)+\ldots+\omega(p_{n(p)})=\omega(k_1)+\ldots+\omega(k_{n(k)}).
\end{equation} 
This indicates that, strictly speaking, the unitary operator $U(h)$ satisfying requirements 
\eqref{Uh}-\eqref{NH} does not exist. However, in calculations of the small-$h$ asymptotic expansions of certain quantities [e.g. the ground state energy $E_{vac}(m,h)$] the  resonance terms do not appear, and the modified FFPT can be effectively used and leads to unambiguous results. This situation is similar to the perturbation theory for  nonlinear systems in classical mechanics \cite{Arnold}. The Birkhoff's theorem states that, if the classical  nonlinear system is close to  some linear one, and {\it the characteristic frequencies of the  latter
do not satisfy the resonance relations}, the dynamics of the nonlinear system can be well approximated
by the integrable system which  Hamiltonian has the Birkhoff  normal form, see page 387 in \cite{Arnold}. The unitary operator $U(h)$ can be viewed as the quantum analogue of the canonical
transform, which maps the original  Hamiltonian of a non-integrable classical system to the 
integrable Birkhoff normal form. 

The second difficulty, which is inherent to the FFPT, comes from the kinematic singularities in the 
matrix elements of the spin order operator between the states with nonzero numbers of kinks. 	 Such
singularities contributing in the leading and higher-orders  of the FFPT lead to infinite and ill-defined quantities like '$\delta(0)$', which require regularization. This problem has been
widely discussed in the literature, mostly in the context of    finite-temperature correlation function  calculations 
 \cite{LeCl99,Saleur2000,Alt06,EsKon09,Tak10}. Several regularization procedures have been proposed, such as  finite 
volume regularization \cite{Tak10,PozTak10}, and appropriate infinitesimal shiftings of the kinematic poles  into the complex plane \cite{LeCl99,EsKon09,Rut09}. Here we apply a different regularization scheme, which seems to be more convenient  for the problem  considered. 

Keeping the length of the system infinite, we replace
the uniform magnetic field $h>0$ by the non-uniform field 
 $h_R({\rm x})$, which is switched on only in the large, but finite  interval  $[-R/2,R/2]$, $R\gg m^{-1}$,
\begin{align}
h_R({\rm x}) = h\,\, \chi( {\rm x};-R/2,R/2),\\
{\rm where}\;\;\chi( {\rm x};-R/2,R/2)=\begin{cases}\nonumber
1, &{\rm if}\; {\rm x}\in [-R/2,R/2],\\
0,&{\rm if}\; {\rm x}\notin [-R/2,R/2].
\end{cases}
\end{align}
After performing all calculations, we proceed to the limit $R\to\infty$.
 
Accordingly, instead of the IFT Hamiltonian \eqref{Ham}, we get a set of Hamiltonians 
$\mathcal{H}_R$ parametrized by the length $R$,
\begin{eqnarray}\label{HR}
&&\mathcal{H}_R=\mathcal{H}_0+h\,V_R,\\
&&V_R=-\int_{-R/2}^{R/2}  d\rm{x}\,\sigma(\rm{x}). \label{VR}
\end{eqnarray}
After diagonalization of the Hamiltonian $\mathcal{H}_R$ in the fermionic number along the lines
described in Section 5 of \cite{Rut09}, we arrive to equations (35)-(39) of \cite{Rut09}, modified by  the following replacements: 
\begin{equation}\label{rep}
V\to V_R, \quad \mathcal{H}\to\mathcal{H}_R,\quad U\to U_R, \quad \Lambda\to \Lambda_R. 
\end{equation}
In the rest of this Section, the efficiency of the described version of the  FFPT will be 
demonstrated by the recovery  of some well-known features of the IFT in the weak 
confinement regime and the derivation of  several new results.
\subsection{Vacuum sector}
To warm-up, let us consider the small-$h$ expansion of the ferromagnetic
ground state energy in the IFT. The results will be used in the subsequent subsection 
in calculations of the radiative corrections to the 
kink dispersion law and string tension. 

The expansion of the ground state energy  $E_{vac}(m,h,R)$ can be read from Subsection~5.1 of  
Reference \cite{Rut09}, with substitutions \eqref{rep}:
\begin{eqnarray} \nonumber
 &&E_{vac}(m,h,R)\equiv \langle  \underline{0}| \mathcal{H}_R| \underline{0}\rangle=\\
&&\langle  \underline{0}| U_R(h)( \underline{\mathcal{H}}_0+h \underline{V}_R)U_R(h)^{-1}| \underline{0}\rangle=
\sum_{j=1}^\infty \delta_{j} E_{vac}(m,h,R) ,\label{RSch}
\end{eqnarray}
where $\delta_{j} E_{vac}(m,h,R)\sim h^j$, and 
\begin{eqnarray} \label{E2vac}
&&\delta_{1}E_{vac}(m,h,R)=h \langle  {0}| {V}_R| {0}\rangle=-h\bar{\sigma}R ,\\
&&\delta_{2}E_{vac}(m,h,R)=-h^2\sum_{q\atop{n(q)\ne 0}} \label{al2}
\frac{\langle  {0} | {V}_R|q\rangle \langle q|  {V}_R| {0}\rangle}
{\omega(q)}, \\
&&\delta_{3}E_{vac}(m,h,R)=-h^3\langle  {0}| {V}_R| {0}\rangle \sum_{q\atop{n(q)\ne 0}}
 \frac{\langle  {0}| {V}_R|q\rangle \langle q| {V}_R| {0}\rangle}
{[\omega(q)]^2}       \label{al3}
 +\\
 &&h^3\sum_{q,q'\atop{n(q)\ne 0\ne n(q')}}\frac{\langle  {0}| {V}_R|q\rangle\,
\langle q| {V}_R| {q'}\rangle
\, \langle q'| {V}_R| {0}\rangle}{\omega(q)\,\omega(q')}
. \nonumber
\end{eqnarray}
The same abbreviation as in equation \eqref{mcF} have been used, 
$n(q)$ denotes the number of fermions in the intermediate state $|q\rangle \equiv | q_1, q_2,\ldots q_{n(q)}\rangle$. 

Four comments on equations \eqref{RSch}-\eqref{al3}  are in order. 
\begin{enumerate}
\item 
There are no resonance poles [like in equation \eqref{mcF}] in expansion $\eqref{RSch}$, while
the kinematic singularities are present in  its third and higher order terms.
\item
Equation \eqref{RSch} is nothing else but the Rayleigh-Schr{\"o}dinger expansion (see, for example 
\textsection 38 in \cite{landau1981quantum}) for the  ground state energy of   the Hamiltonian \eqref{HR}. This expansion in $h$ is asymptotic. In the limit $R\to\infty$, its convergence radius  goes 
to zero due  to the weak 
essential {\it droplet singularity} \cite{FonZam2003,LANGER1967108,Rut99} at $h=0$ of the IFT ground state energy density $\rho(m,h)$. The latter can be identified with  the limit
\begin{equation}\label{rhn}
\rho(m,h)\equiv\lim_{R\to\infty} \frac{E_{vac}(m,h,R)}{R}=\sum_{j=1}^\infty \delta_{j} \rho(m,h),  
\end{equation}
where $\delta_{j} \rho(m,h)\sim h^j$.
\item The ground state energy density $\rho(m,h)$ is simply related to the universal function 
${F}(m,h)$ that describes the singular part of the free  energy in the vicinity of the 
critical point in the two-dimensional Ising model universality class \cite{FonZam2003,Bazh10},
\begin{equation}
\rho(m,h)={F}(m,h)-{F}(m,0)
=m^2 \,G_{\rm low}(\xi),
\end{equation}
where $\xi=h/|m|^{15/8}$, and the zero-field term ${F}(m,0)$ describes Onsager's singularity 
\cite{Ons44} of the Ising free energy at zero $h$, 
\begin{equation}
{F}(m,0)=\frac{m^2}{8\pi} \ln m^2.
\end{equation}
The scaling function $G_{\rm low}(\xi)$ can be expanded into the asymptotic expansion in powers of 
$\xi$ 
\begin{equation}\label{Gt}
G_{\rm low}(\xi)\simeq \tilde{G}_1 \xi+ \tilde{G}_2 \xi^2+\ldots,
\end{equation}
whose initial coefficients are known with  high accuracy \cite{McCoyWu78,FonZam2003,Bazh10}.
\item Fonseca and Zamolodchikov argued \cite{FZ06}, that the perturbation expansion
for the renormalized string tension $f(m,h)$, which characterizes the linear attractive potential acting between two kinks 
at large  distances, is related with  the ground state energy density $\rho(m,h)$ in the following way
\begin{equation} \label{fh}
f(m,h)={\rho(m,-h)-\rho(m,h)},
\end{equation}
where the right-hand side is understood in the sense of the formal perturbative 
expansion in $h$. 
Combining \eqref{rhn} and \eqref{fh}, we get
\begin{equation}\label{fh1}
f(m,h)=\sum_{j=0}^\infty f^{(2j+1)}(m,h),
\end{equation}
where
\begin{equation}\label{fhj}
 f^{(2j+1)}(m,h)=-2
\lim_{R\to\infty} \frac{\delta_{2j+1}E(m,h,R)}{R}\sim h^{2j+1}, 
\end{equation}
and 
\begin{equation}
f^{(1)}(m,h)=f_0(h)\equiv 2 h \bar{\sigma}.
\end{equation}
\end{enumerate}

The second-order term  $\delta_{2}E_{vac}(m,h,R)$ is defined by means of the  Lehmann expansion \eqref{al2}, whose  explicit form reads as
\begin{equation}\label{dE2a}
\delta_{2}E_{vac}(m,h,R)=\sum_{l=1}^\infty \delta_{2,2l}E_{vac}(m,h,R),
\end{equation}
where
\begin{eqnarray}\label{a22L}
\delta_{2,\nu}E_{vac}(m,h,R)=-\frac{h^2}{\nu!}\iint_{-R/2}^{R/2}d{\rm x}_1\,d{\rm x}_2 
\int_{-\infty}^\infty \frac{dq_1\ldots dq_\nu}{(2\pi)^\nu}\cdot \\
\frac{\exp[i(q_1+\ldots+q_\nu)({\rm x_1}-{\rm x_2})]}{\omega(q_1)+\ldots+\omega(q_\nu)}\,
\langle 0|\sigma(0)|q_1,\ldots,q_\nu \rangle\langle q_\nu,\ldots,q_1|\sigma(0)|0\rangle.\nonumber
\end{eqnarray}
Straightforward summation of  (\ref{dE2a}) yields,
\begin{equation} \label{dE2b}
\delta_{2}E_{vac}(m,h,R)=-h^2\iint_{-R/2}^{R/2}d{\rm x}_1\,d{\rm x}_2 \int_{0}^\infty d{\rm y}_1 \,
\langle 0 |\Delta \sigma({\rm x}_1,{\rm y}_1)\,\Delta \sigma({\rm x}_2,0)|0\rangle,
\end{equation}
where
\[
\Delta \sigma({\rm x},{\rm y})=\exp( \mathcal{H}_0\, {\rm y} )
\sigma({\rm x})\exp(- \mathcal{H}_0\, {\rm y} )- \bar{\sigma}.
\]
Since the matrix element in the integrand does not depend on  $({\rm x}_1+{\rm x}_2)/2$
and  vanishes exponentially for $|{\rm x}_1-{\rm x}_2|\, m \gg1$, we can easily proceed  
to the limit $R\to\infty$ in 
\eqref{dE2b}, arriving at 
the well-known representation of the magnetic susceptibility 
in terms of the spin-spin correlation function,
\begin{eqnarray}
\delta_2 \rho(m,h)\equiv \lim_{R\to\infty} \frac{\delta_{2,\nu}E_{vac}(m,h,R)}{R}= \\
-h^2 \int_{-\infty}^\infty d{\rm x}\int_{0}^\infty d{\rm y} \,
\langle 0 |\Delta \sigma({\rm x},{\rm y})\,\Delta \sigma(0,0)|0\rangle.\nonumber
\end{eqnarray}

Let us return now to the Lehmann expansion \eqref{dE2a} for the ground state energy, 
perform the  elementary integration 
over ${\rm x}_1,\,{\rm x}_2$ in \eqref{a22L}, and proceed to the limit $R\to\infty$, exploiting the equality
\begin{equation}
\lim_{R\to\infty}\frac{4  \sin^2 ( q R/2)}{R\, q^2} = 2\pi \delta(q).
\end{equation}
As a result, we arrive at  the familiar  spectral expansion \cite{McCoy76} for the 
ground state energy density
\begin{eqnarray}\label{ro2}
&&\delta_2 \rho(m,h)=\sum_{l=1}^\infty \delta_{2,2l}\, \rho(m,h),\\
&&\delta_{2,2l}\, \rho(m,h)=-h^2\,\frac{1}{(2 l)^2}\int_{-\infty}^\infty 
\frac{dq_1\ldots dq_{2l}}
{(2\pi)^{2l-1}}
\frac{\delta(q_1+\ldots+q_{2l})}{\omega(q_1)+\ldots+\omega(q_\nu)}\cdot \\
&&\langle 0|\sigma(0)|q_1,\ldots,q_\nu \rangle\langle q_\nu,\ldots,q_1|\sigma(0)|0\rangle.\nonumber
\end{eqnarray}
The first term in expansion \eqref{ro2} can be easily calculated using the explicit expressions 
 \eqref{fIs2}, \eqref{fIs3} for the
 form factors, giving
\begin{equation}
\delta_{2,2} \rho(m,h)=-\frac{h^2\bar{\sigma}^2}{12\pi m}.
\end{equation}
The corresponding two-fermion contribution $\tilde{G}_{2,2}$ to the universal amplitude $\tilde{G}_2$ 
\begin{equation}\label{G22}
\tilde{G}_{2,2}=-\frac{{\bar{s}}^2}{12\pi}=-0.0489063\ldots
\end{equation}
reproduces the well-known result of Tracy and McCoy \cite{Tracy73},
which is rather close to the exact value \cite{McCoy76,FonZam2003,Bazh10}  $\tilde{G}_2=-0.0489532897203\ldots$

Now let us turn to the third order term \eqref{al3} in the expansion \eqref{RSch} for the ground state 
energy $E_{vac}(m,h,R)$. Unlike the previous case of the second-order correction,
 kinematic singularities do contribute to $\delta_{3}E_{vac}(m,h,R)$ through  the 
 matrix element $\langle q| {V}_R| {q'}\rangle$ in the second line 
 of \eqref{al3}. Nevertheless,    the right-hand side of  \eqref{al3} 
is well defined due to the chosen regularization \eqref{rep}. 

After summation of the 
Lehmann expansion in \eqref{al3} one arrives 
in the  limit $R\to\infty$ at 
the well-known integral representation \cite{McCoy78} for $\delta_3 \rho(m,h)$
in terms of the three-point correlation function,
\begin{eqnarray}\label{d3Rho}
&&\delta_3 \rho(m,h)=\\
&&-h^3 \iint_{-\infty}^\infty d{\rm x}_1d{\rm x}_3\int_{0}^\infty d{\rm y}_1 
\int_{-\infty}^0 d{\rm y}_3\,
\langle 0 |\Delta \sigma({\rm x}_1,{\rm y}_1)\Delta \sigma(0,0)
\Delta \sigma({\rm x}_3,{\rm y}_3)|0\rangle.\nonumber
\end{eqnarray}

Alternatively, one can truncate the spectral series  \eqref{al3} which defines \newline
$\delta_3 \,E_{vac}(m,h,R)$ 
 at the level of the 
two-kink intermediate states $n(q)=n(q')=2$. Denoting the result by 
$\delta_{3,2} \,E_{vac}(m,h,R)$, we get explicitly 
\begin{equation}\label{Evac32}
\delta_{3,2} \,E_{vac}(m,h,R)=A_{3,2}(m,h,R)+B_{3,2}(m,h,R),
\end{equation}
where
\begin{eqnarray}\label{dE3}
&&A_{3,2}(m,h,R)=\frac{h^3\bar{\sigma} R}{2} \iint_{-\infty}^\infty
\frac{dq_1dq_2}{(2\pi)^2}\frac{1}{[\omega(q_1)+\omega(q_2)]^2}\cdot \\\nonumber
&&\iint_{-R/2}^{R/2} d{\rm x}_1 d{\rm x}_2 \,e^{i({\rm x}_1 -{\rm x}_2)(q_1+q_2)}
\langle0|\sigma(0)|q_1,q_2\rangle \langle q_2,q_1|\sigma(0)|0\rangle, \\\label{Bint}
&&B_{3,2}(m,h,R)=-\frac{h^3}{4} \iint_{-\infty}^\infty\frac{dq_1dq_2}{(2\pi)^2}   
\frac{1}{[\omega(q_1)+\omega(q_2)]}\iint_{-\infty}^\infty\frac{dq_1'dq_2'}{(2\pi)^2} \cdot\\\nonumber
&&
\frac{1}{[\omega(q_1')+\omega(q_2')]}\iiint_{-R/2}^{R/2} d{\rm x}_1 d{\rm x}_2 d{\rm x}_3  \,e^{i({\rm x}_1 -{\rm x}_2)(q_1+q_2)}
 \,e^{i({\rm x}_2 -{\rm x}_3)(q_1'+q_2')}\cdot \\\nonumber
 &&\langle0|\sigma(0)|q_1,q_2\rangle  \langle q_2,q_1|\sigma(0)|q_1',q_2'\rangle
 \langle q_2',q_1'|\sigma(0)|0\rangle.
\end{eqnarray} 
Here the two-kink matrix elements of the spin operator  are determined by 
equations \eqref{fIs}-\eqref{fIs3}, while the
four-kink matrix element in the last line can be expressed in terms of the latter by means of the Wick expansion:
\begin{eqnarray}
 \langle q_2,q_1|\sigma(0)|q_1',q_2'\rangle= [\langle q_2,q_1|\sigma(0)|0\rangle \langle 0|\sigma(0)|q_1',q_2'\rangle+ \\
 \langle q_1|\sigma(0)|q_1'\rangle \langle q_2|\sigma(0)|q_2'\rangle-
  \langle q_1|\sigma(0)|q_2'\rangle \langle q_2|\sigma(0)|q_1'\rangle] \bar{\sigma}^{-1}.\nonumber
\end{eqnarray}
Since the two last terms in the square brackets in the right-hand side provide equal contributions to the  
integral  \eqref{Bint}, we can replace the four-kink matrix element in its
integrand   as follows
\begin{equation}\label{2si}
 \langle q_2,q_1|\sigma(0)|q_1',q_2'\rangle\leadsto [\langle q_2,q_1|\sigma(0)|0\rangle \langle 0|\sigma(0)|q_1',q_2'\rangle+ 2
 \langle q_1|\sigma(0)|q_1'\rangle \langle q_2|\sigma(0)|q_2'\rangle] \bar{\sigma}^{-1}.
\end{equation}
The second term in  the bracket containing the product of two kinematic singularities 
can be modified to the 
form 
\begin{eqnarray}\label{lr}
 &&2\langle q_1|\sigma(0)|q_1'\rangle \langle q_2|\sigma(0)|q_2'\rangle=\\
 &&- 2\,\bar{\sigma}^2\,
 \frac{\omega(q_1)+\omega(q_1')}{\sqrt{\omega(q_1) \omega(q_1')}}\,
 \frac{\omega(q_2)+\omega(q_2')}{\sqrt{\omega(q_2) \omega(q_2')}}\,
 \mathcal{P}\,\frac{1}{q_1-q_1'} \mathcal{P}\,\frac{1}{q_2-q_2'}=\nonumber \\
 && 8\pi^2 \bar{\sigma}^2\,\delta(q_1-q_1')\,\delta(q_2-q_2')-
{\bar{\sigma}^2}\, \frac{\omega(q_1)+\omega(q_1')}{\sqrt{\omega(q_1) \omega(q_1')}}
 \frac{\omega(q_2)+\omega(q_2')}{\sqrt{\omega(q_2) \omega(q_2')}}\cdot \nonumber\\
 &&\left(
 \frac{1}{q_1-q_1'+i0} \,\frac{1}{q_2-q_2'-i0}+ \frac{1}{q_1-q_1'-i0} \,\frac{1}{q_2-q_2'+i0}
 \right).\nonumber
 \end{eqnarray}
In deriving \eqref{lr} we have used \eqref{fIs}, \eqref{Cau}, together with the  equality
 \begin{eqnarray}
 \mathcal{P}\,\frac{1}{q_1-q_1'} \mathcal{P}\,\frac{1}{q_2-q_2'}=-\pi^2\,\delta(q_1-q_1')\,\delta(q_2-q_2')+\\
 \frac{1}{2}\left(
 \frac{1}{q_1-q_1'+i0} \,\frac{1}{q_2-q_2'-i0}+ \frac{1}{q_1-q_1'-i0} \,\frac{1}{q_2-q_2'+i0}
 \right).\nonumber
\end{eqnarray}
After substitution of \eqref{lr} into \eqref{2si},  \eqref{Bint}, the term  
$
8\pi^2 \bar{\sigma}^2\,\delta(q_1-q_1')\,\delta(q_2-q_2')
$
 in the right-hand side of \eqref{lr}
 gives rise to the contribution in $B_{3,2}(m,h,R)$, which cancels exactly with the term $A_{3,2}(m,h,R)$
 in \eqref{Evac32}.  Performing the integration over ${\rm  x}_1, {\rm  x}_2, {\rm  x}_3$ over the cube $(-R/2,R/2)^3$ 
 in the remaining part and 
 dividing the result by  $R$, we obtain
\begin{eqnarray}\nonumber
&&\frac{\delta_{3,2} \,E_{vac}(m,h,R)}{R}=-\frac{h^3\bar{\sigma}^3}{4}\int_{-\infty}^\infty
\frac{dq_1dq_2dq_1'dq_2'}{(2 \pi)^4}\,\Delta_3(q_1+q_2,q_1'+q_2',R) \cdot\\
&&\mathcal{G}(q_1,q_2,q_1',q_2'),  \label{dEa}
 \end{eqnarray} 
 where
\begin{equation} \label{Del}
 \Delta_3(p,k,R)=\frac{8\sin (p R/2)\,\sin (k R/2)\,\sin[ (k-p) R/2]}{R\,p\, k\, (k-p)},
\end{equation} 
 and 
 \begin{eqnarray}
\mathcal{G}(q_1,q_2,q_1',q_2')= \frac{\omega(q_1)-\omega(q_2)}{\sqrt{\omega(q_1)\, \omega(q_2)}}\,
\frac{\omega(q_2')-\omega(q_1')}{\sqrt{\omega(q_1')\, \omega(q_2')}}
\frac{1}{(q_1+q_2)(q_1'+q_2')}\cdot\\\nonumber
 \frac{1}{\omega(q_1)+\omega(q_2)} \frac{1}{\omega(q_1')+\omega(q_2')}\Bigg\{
 \frac{\omega(q_1)-\omega(q_2')}{\sqrt{\omega(q_1') \omega(q_2')}}\,
 \frac{\omega(q_2)-\omega(q_1)}{\sqrt{\omega(q_1) \,\omega(q_2)}}\cdot \\
 \frac{1}{(q_1'+q_2')(q_1+q_2)}+\nonumber
 \frac{\omega(q_1)+\omega(q_1')}{\sqrt{\omega(q_1)\, \omega(q_1')}}\, \frac{\omega(q_2)+\omega(q_2')}{\sqrt{\omega(q_2) \,\omega(q_2')}}\cdot\\\nonumber
\left(
 \frac{1}{q_1-q_1'+i0} \,\frac{1}{q_2-q_2'-i0}+ \frac{1}{q_1-q_1'-i0} \,\frac{1}{q_2-q_2'+i0}
 \right)
\Bigg\}.
 \end{eqnarray} 
It is possible to show that the weak large-$R$ limit of the  function $\Delta_3(p,k,R)$ is proportional 
 to the  two-dimensional 
 $\delta$-function,
 \begin{equation}\label{delt}
\lim_{R\to\infty } \Delta_3(p,k,R)=4\pi^2\,\delta(p)\,\delta(k).
\end{equation} 
The simplest way to prove this  equality
is to integrate $\Delta_3(p,k,R)$ multiplied with the plane-wave
test function. The result reads as
 \begin{equation}\label{intJ}
\iint_{-\infty}^\infty dp\, dk\, \Delta_3(p,k,R)\,\exp[i(p{\rm x}+k {\rm y})]=4\pi^2\,
\left[1-
\frac{\max(|{\rm x}|,|{\rm y}|,|{\rm x+y}|)}{R}\right],
\end{equation} 
if $\max(|{\rm x}|,|{\rm y}|,|{\rm x+y}|)<R$. 
Taking  the limit $R\to\infty$ in \eqref{intJ} , we arrive at \eqref{delt}. 

Exploiting  \eqref{delt}, one can proceed to the limit $R\to\infty$  in \eqref{dEa}, yielding
 \begin{eqnarray}\label{d32ro}
\delta_{3,2} \,\rho(m,h) \equiv \lim_{R\to\infty }  \frac{\delta_{3,2} \,E_{vac}(m,h,R)}{R}=\\
-\frac{h^3\bar{\sigma}^3}{4}\int_{-\infty}^\infty
\frac{dq_1dq_1'}{(2 \pi)^2}\, \mathcal{G}(q_1,-q_1,q_1',-q_1') =
\frac{h^3\bar{\sigma}^3}{16\pi^2\, m^4}\,( C_1+C_2),\nonumber
\end{eqnarray}
where
\begin{equation}\label{I1}
C_1=-\frac{m^4}{4}\left\{\int_{-\infty}^\infty {dq}\,
\frac{q^2}{[\omega(q)]^5}
\right\}^2=-\frac{1}{9},
\end{equation}
and 
\begin{eqnarray}\label{I2}
C_2=-\frac{m^4}{4} \iint_{-\infty}^\infty dq \,dq'\,
\frac{q \,q'\,[\omega(q)+ \omega(q')]^2}{[\omega(q) \omega(q')]^4}\cdot \\
\left[
\frac{1}{(q-q'+i0)^2}+\frac{1}{(q-q'-i0)^2}
\right]=
\frac{4}{3}+\frac{\pi^2}{8}
.
\nonumber
\end{eqnarray}
Calculation of the integral in equation \eqref{I1} is straightforward. 
The calculation of the double 
integral $C_2$  is harder and described in \ref{int2}.

Combining \eqref{d32ro}-\eqref{I2}, we obtain finally
\begin{eqnarray}
\delta_{3,2} \rho(m,h)=\frac{h^3\bar{\sigma}^3}{16 m^4}\left(
\frac{11}{9\pi^2}+\frac{1}{8}
\right).
\end{eqnarray}
For the two-kink contribution $\tilde{G}_{3,2}$ to the amplitude 
$\tilde{G}_3$, this yields
\begin{eqnarray}\label{G32}
\tilde{G}_{3,2}=\frac{\bar{s}^3}{16 }\left(
\frac{11}{9\pi^2}+\frac{1}{8}
\right)=0.0389349\ldots
\end{eqnarray}

The exact value of the universal amplitude $\tilde{G}_3$ is unknown. 
In 1978, 
McCoy and Wu \cite{McCoyWu78} performed a thorough analysis of the three- and four-point
 spin correlation functions in the zero-field 
Ising model on the square lattice, from which they obtained the approximate value for this amplitude,
\begin{equation}\label{GMc}
\tilde{G}_3\approx \frac{11 \bar{s}^3}{72}=0.0387529\ldots
\end{equation}

Recently, at least six digits of the exact amplitude $\tilde{G}_3$ have become available
\begin{equation}\label{G3}
\tilde{G}_3=0.0388639\ldots
\end{equation}
due to the very accurate numerical calculations carried out by 
Mangazeev {\it et al.} \cite{Mang09,Bazh10} 
for the square and triangular lattice Ising models. \begin{footnote}
{The values of the amplitude $\tilde{G}_3$ reported in \cite{Bazh10} 
for the square and triangular lattices are
0.038863932(3) and 0.0388639290(1), respectively. }
\end{footnote}

Comparison of \eqref{G32} and  \eqref{GMc} with \eqref{G3} indicates, that 
(i) the two-kink contribution \eqref{G32} approximates the "exact"  amplitude \eqref{G3} 
somewhat better than \eqref{GMc}; 
(ii) the two-kink configurations provide the dominant contribution to the universal amplitude $\tilde{G}_3$.The configurations
with four and more kinks in  intermediate states contribute less then $0.2\%$ in the spectral sum \eqref{al3}.
\subsection{One-fermion sector \label{1FS}}
In this subsection we address the modified FFPT in the one-fermion sector $n(\underline{p})=n(\underline{k})=1$, and  extend it to the third order in $h$. 

The matrix element of the Hamiltonian \eqref{HR} between the dressed one-fermion states  $\langle\underline{p}|$ and 
$|\underline{k}\rangle$ can be written as
\begin{eqnarray} \label{1sect}
 \langle \underline{p}|\mathcal{H}_R|\underline{k}\rangle =
  \langle {p}|U_R(h)\,\mathcal{H}_R\,U_R(h)^{-1}|{k}\rangle=
 2\pi \delta(p-k)\,\omega(p)+
 \delta\langle \underline{p}|\mathcal{H}_R|\underline{k}\rangle.
 \end{eqnarray}
 Expanding here the unitary operator $U_R(h)$ and its inverse in  powers 
 of $h$, one arrives at the perturbation expansion 
 \begin{equation}\label{delta2}
 \delta\langle \underline{p}|\mathcal{H}_R|\underline{k}\rangle= 
 \sum_{j=1}^\infty \delta_j \langle \underline{p}|\mathcal{H}_R|\underline{k}\rangle.
 \end{equation}
 Three initial terms in this expansion can be obtained from  equation (37)-(39) of \cite{Rut09}
by means of the replacements  \eqref{rep}:
\begin{eqnarray} 
 \label{d21}
&&\delta_1 \langle \underline{p}|\mathcal{H}_R|\underline{k}\rangle=h  \langle p| {V}_R|k\rangle,
\\\label{d22}
&& \delta_2 \langle \underline{p}|\mathcal{H}_R|\underline{k}\rangle=  -
\frac{h^2}{2}\!\!\!\!\sum_{q\atop {n(q)\ne n({p})}}\!\!\!\langle p | {V}_R|q\rangle \langle q|  {V}_R|k\rangle 
\left\{\frac{ 1 }{\omega(q)- \omega(p)} +\frac{ 1 }{\omega(q)- \omega(k)}\right\}\!\!,\\
&& \delta_3 \langle \underline{p}|\mathcal{H}_R|\underline{k}\rangle=   \nonumber 
\frac{h^3}{2}\sum_{q,q'}  
\langle p | {V}_R|q\rangle \langle q|  {V}_R|q'\rangle\langle q'|  {V}_R|k\rangle \Bigg\{ 
[1-\delta_{n(q),n({p})}][1-\delta_{n(q'),n({p})}]\\\label{delta3}
&& \cdot\Bigg[\frac{1}{[\omega(p)-\omega(q)]}\frac{1}{[\omega(p)-\omega(q')]}  
 +\frac{1}{[\omega(k)-\omega(q)]}\frac{1}{[\omega(k)-\omega(q')]}\Bigg] \\ 
&&+ \frac{1}{\omega(q)-\omega(q')}\left[\frac{\delta_{n(q),n(\underline{p})}
 [1-\delta_{n(q'),n({p})}]}{\omega(q')-\omega(p)} - 
\frac{[1-\delta_{n(q),n({p})}]\delta_{n(q'),n({p})}}{\omega(q)-\omega(k)}\right]\Bigg\},   \nonumber
\end{eqnarray}
where $n(p)=n(k)=1$. 

One can easily see, that the matrix elements 
$\delta_j \langle \underline{p}|\mathcal{H}_R|\underline{k}\rangle$
obey the following symmetry relations:
\begin{eqnarray}\label{hh}
\delta_j \langle \underline{p}|\mathcal{H}_R|\underline{k}\rangle&=&
(-1)^j\,\delta_j \langle \underline{k}|\mathcal{H}_R|\underline{p}\rangle,\\
\delta_j \langle \underline{p}|\mathcal{H}_R|\underline{k}\rangle&=&(-1)^j\,[\delta_j 
\langle \underline{p}|\mathcal{H}_R|\underline{k}\rangle]^*,
\end{eqnarray}
for $j=1,2,\ldots$
The kinematic singularity is present already in the first order term
\eqref{d21}. The resonance poles contribute to the second and higher orders of expansion
\eqref{delta2} for   large enough momenta $p$ and $k$, due to the terms, like those
in braces in \eqref{d22}, \eqref{delta3}. 
Nevertheless, at finite $R$, the right-hand sides of equations \eqref{d21}-\eqref{delta3} determine 
well defined generalized functions, if the absolute values of momenta $p$ and $k$ are small enough,  
\begin{equation}\label{pkm}
\omega(p)< 3m, \quad {\rm and}\quad \omega(k)< 3m. 
\end{equation}
The latter conditions guarantee that the resonance poles do not appear in expansion
\eqref{delta2}.
The constrains \eqref{pkm} will be imposed in the subsequent FFPT calculations at finite $R$. 
After proceeding to the limit $R\to \infty$, the results will be analytically continued to 
larger momenta, $|p|>\sqrt{2} \,m$.

We postulate the following definition of the renormalized quark dispersion law $\epsilon(p,m,h)$,
\begin{eqnarray}\label{eps}
\lim_{R\to\infty}\left\{ \langle \underline{p}|\mathcal{H}_R|\underline{k}\rangle
-\pi \delta(p-k)\left[
E_{vac}(m,h,R)+E_{vac}(m,-h,R)
\right]
\right\}=\\
2\pi\, \epsilon(p,m,h)\,\delta(p-k)+2\pi i f(m,h)\, \delta'(p-k).\nonumber
\end{eqnarray}
Just as in the case of definition \eqref{fh},  both sides in the above equation must be understood as  formal power series in $h$. Equating the coefficients in these power series and taking into account \eqref{hh} and \eqref{fhj}, one  finds
\begin{equation}\label{epev}
\lim_{R\to\infty}\left\{\delta_{j} \langle \underline{p}|\mathcal{H}_R|\underline{k}\rangle
-2\pi \delta(p-k)\,
\delta_{j} \langle \underline{0}|\mathcal{H}_R|\underline{0}\rangle
\right\}=2\pi\, \delta_{j}\,\epsilon(p,m,h)\,\delta(p-k),
\end{equation}
for even $j=2,4,\ldots$, and 
\begin{eqnarray}\label{Hod}
&&\lim_{R\to\infty}\left\{\delta_{j} \langle \underline{p}|\mathcal{H}_R|\underline{k}\rangle
+4\pi i \,\delta'(p-k) R^{-1} \,\delta_{j} \, \langle \underline{0}|\mathcal{H}_R|\underline{0}\rangle
\right\}=0,\\
&&\delta_{j}\,\epsilon(p,m,h)=0,\label{epsod}
\end{eqnarray}
for odd $j=1,3,\ldots$
So, we can argue on the basis of the  above heuristic  analysis, that  the Taylor expansion of the quark dispersion law $\epsilon(p,m,h)$ contains only even powers
of $h$, which are determined by equation \eqref{epev}. 

It was shown in \cite{FZWard03} that 
 the renormalized quark dispersion law $\epsilon(p,h)$, 
does not have the Lorentz covariant form  in the confinement regime. 
Nevertheless, 
the 'dressed quark mass' $m_q(m,h)$
can be extracted from large-$p$ asymptotics of $\epsilon(p,h)$
in the following way \cite{FZWard03, Rut09},
\begin{equation}\label{mq2}
[m_q(m,h)]^2 =\lim_{p\to\infty}\{2 p\, [\epsilon(p,m,h)-p]\}.
\end{equation}
This relation is understood, of course,  in the sense  of a 
 power series in $h$, or, equivalently,  in the parameter $\lambda= 2 h\bar{\sigma}/m^2$. It follows from
 \eqref{epsod}, that this expansion contains only even powers,
\begin{equation}\label{mqS}
{m_q^2} =m^2+m^2 \sum_{l=1}^\infty a_{2l} \lambda^{2l}.
\end{equation}
In order to validate the latter statement, it remains to  show that the large-$R$ limits in the left-hand sides of equations 
\eqref{epev}  and \eqref{Hod} exist, and to prove equalities \eqref{Hod}. In what follows, we shall do it for the three 
initial values  $j=1,2,3$.

The case $j=1$ is  quite simple.
The term (\ref{d21})  linear in $h$  in  expansion \eqref{delta2}  reads as
\begin{eqnarray}\label{V1}
\delta_1 \langle \underline{p}|\mathcal{H}_R|\underline{k}\rangle =-h\int_{-R/2}^{R/2} d {\rm x}\,  \langle p|\sigma({\rm x})|k\rangle=\\
  i\,h\bar{\sigma}\,\frac{ \omega(p)+\omega(k) }{[\omega(p)\omega(k)]^{1/2}}\,
 \frac{2 \sin[R(k-p)/2]}{(k-p)}\,
\mathcal{P} \frac{ 1 }{k-p}.\nonumber
\end{eqnarray}
Even though the right-hand side contains the kinematic singularity, it describes a well defined
generalized function at arbitrary finite $R$. Furthermore, exploiting the equality
\begin{equation}\label{R1}
  \lim_{R\to\infty} \frac{2\sin(q R/2)}{q}\,\mathcal{P}\,\frac{1}{q}=-2\pi \delta'(q),
\end{equation}   
we can proceed to the limit $R\to\infty$ in equation \eqref{V1},  obtaining
 \begin{equation}  \label{R2}
\lim_{R\to\infty}\delta_1  \langle \underline{p}|\mathcal{H}_R|\underline{k}\rangle   = 4\pi i\,
\delta'(p-k)\,  h\bar{\sigma}.
\end{equation}     
This proves  \eqref{Hod} for $j=1$, since 
$\delta_1\langle \underline{0}|\mathcal{H}_R|\underline{0}\rangle=-h \bar{\sigma}R$.

Turning to the term \eqref{d22} quadratic in $h$, we first perform the summation over the 
number $n(q)$ of the fermions in the intermediate state $|q\rangle$,  
subject to the  requirement  \eqref{pkm}. The result can be written in the 
compact form
 \begin{eqnarray}\label{d2pk}
&&\delta_2 \langle \underline{p}|\mathcal{H}_R|\underline{k}\rangle =
-\frac{h^2}{2} \int_{0}^\infty d{\rm y} \,
\iint_{-R/2}^{R/2}d{\rm x}_1\,d{\rm x}_2 \,\left(1+e^{y[\omega(k)-\omega(p)]}\right)\cdot \\
&&\langle p | \sigma({\rm x}_1-{\rm x}_2,{\rm y})(1-P_1) \sigma(0,0)|k\rangle
e^{i {\rm x}_2(k-p)}
,\nonumber
\end{eqnarray}
where ${P}_1$ denotes the orthogonal projection operator onto the 
one-fermion subspace of the Fock space. 
The matrix element in the right-hand side can be represented as 
 \begin{eqnarray}\nonumber
&&\langle p | \sigma({\rm x},{\rm y})(1-P_1) \sigma(0,0)|k\rangle=
2\pi\, \delta(p-k)[\langle 0 | \sigma({\rm x},{\rm y})
 \sigma(0,0)|0\rangle-\bar{\sigma}^2]+\\
&&\langle p | \sigma({\rm x},{\rm y})(1-P_1) \sigma(0,0)|k\rangle_{reg},
 \label{pk}
\end{eqnarray}
where ${\rm x}={\rm x}_1-{\rm x}_2$.
The first singular term in the right-hand side 
represents the 'direct propagation part' \cite{FZWard03}, while the second term is a regular function
of momenta  at $k\to p$. 

After substitution of  \eqref{pk} into  \eqref{d2pk}
and subtraction  the singular term 
we get 
 \begin{eqnarray}\nonumber
&&\delta_2 \langle \underline{p}|\mathcal{H}_R|\underline{k}\rangle -2\pi \delta(p-k)\,
\delta_2 E_{vac}(h,R)=\\\label{DE2}
&&-\frac{h^2}{2} \int_{0}^\infty d{\rm y} \,
\iint_{-R/2}^{R/2}d{\rm x}_1\,d{\rm x}_2 \,\left(1+e^{y[\omega(k)-\omega(p)]}\right)
e^{i {\rm x}_2(k-p)}
\cdot \\\nonumber
&&\langle p | \sigma({\rm x}_1-{\rm x}_2,{\rm y})(1-P_1) \sigma(0,0)|k\rangle_{reg}.
\nonumber
\end{eqnarray}
In this equation we can safely proceed to the limit $R\to\infty$. 
Comparing  the 
result with  \eqref{eps}, one finds
the second order correction to the kink dispersion law
\begin{eqnarray}\label{d2eps}
&&\delta_2\, \epsilon(p,m,h)=
-{h^2} \int_{0}^\infty d{\rm y} \,
\int_{-\infty}^{\infty}d{\rm x} \cdot\\
&&\lim_{k\to p}\left[\langle p | \sigma({\rm x},{\rm y}) \sigma(0,0)|k\rangle-
\langle p | \sigma({\rm x},{\rm y}) {P}_1 \sigma(0,0)|k\rangle
\right].\nonumber
\end{eqnarray}
Even though the above relation was derived for small  $|p|$ satisfying the first inequality
in \eqref{pkm}, we shall extend it to all real momenta $p$ by analytic continuation. 

The second order correction to the squared quark mass can be read from 
\eqref{mq2} and \eqref{d2eps},
\begin{eqnarray}\label{d2mq}
&&\delta_2 \,[m_q(m,h)]^2=
-2{h^2} \lim_{\beta\to\infty} \int_{0}^\infty d{\rm y} \,
\int_{-\infty}^{\infty}d{\rm x}\cdot\\
&&\lim_{\beta'\to \beta}\left[\langle \beta' | \sigma({\rm x},{\rm y}) \sigma(0,0)|\beta\rangle-
\langle \beta' | \sigma({\rm x},{\rm y}) {P}_1 \sigma(0,0)|\beta\rangle
\right].\nonumber
\end{eqnarray}

This integral representation  for the second order correction to the quark mass [written
in a slightly different form \eqref{a2B}] was first 
derived by Fonseca and Zamolodchikov \cite{FZWard03}. Exploiting the Ward identities, they
 managed to express the matrix element
 in the right-hand side in terms of solutions of certain differential equations, and
 obtained the value 
 \begin{equation}\label{aq}
 a_q=\bar{s}^2\cdot 0.142021619(1)\ldots
 \end{equation}
 for the parameter $a_q$,
 \begin{equation} \label{aq1}
 a_q =2 \bar{s}^2\,a_2
  \end{equation}
 by numerical integration of the 
double  integral in \eqref{d2mq}  over the half-plane in polar coordinates $r, \theta$.

It turns out, that the integral in the polar angle   can be evaluated analytically. 
The details of this calculations 
are relegated to \ref{IPA}. The results read as,
\begin{eqnarray}\label{sigth}
&&\mathcal{U}(r)\equiv \int_0^\pi \frac{d\theta}{\pi}\, 
\lim_{\beta'\to \beta}\langle \beta' | \sigma(r \cos\theta ,r \sin\theta) \sigma(0,0)|\beta\rangle
=\\\nonumber
 &&e^{\chi/2}
\bigg\{ r^2  {\mathfrak b}_0'\,\varphi'\, \cosh\frac{\varphi}{2} + \\
&&{\mathfrak b}_0\left[ \sinh\frac{\varphi}{2} +
r\, \varphi'\,\cosh\frac{\varphi}{2}-\frac{ r^2}{4}\left(\sinh\frac{3\varphi}{2}+\sinh\frac{5\varphi}{2}
\right)\right]\bigg\},   \nonumber
\end{eqnarray}
and
\begin{eqnarray}\label{Sth}
&&\mathcal{W}(r)\equiv \lim_{\beta\to \infty}\int_0^\pi  \frac{d\theta}{\pi}\, 
\lim_{\beta'\to \beta}\langle \beta' | \sigma(r \cos\theta ,r \sin\theta)P_1 \sigma(0,0)|\beta\rangle= \\
&&\frac{2 \bar{s}^2}{\pi}
\bigg\{\left[1-2r^2\right] I_0(r) K_0(r)-
2 r \,K_1(r)\left[I_0(r)+r\, I_1(r)
\right]
\bigg\},\nonumber
\end{eqnarray}
where  
${\mathfrak b}_0(r)$ stands for the solution of the second order differential equation
\begin{equation}\label{difb0}
{\mathfrak b}_0''(r)+r^{-1}\, {\mathfrak b}_0(r)= \cosh[2\varphi(r)]\, {\mathfrak b}_0(r),
\end{equation}
which vanishes at $r\to \infty$, and behaves at small $r\to0$ as 
\begin{equation}
{\mathfrak b}_0(r)=\frac{1}{\Omega(r)}+O(r^4).
\end{equation}
The auxiliary functions   $\varphi(r)$, $\chi(r)$, and $\Omega(r)$ were defined in \cite{FZWard03},  
 $I_j(r)$ and $K_j(r)$ are the Bessel function of the imaginary argument and 
the McDonald's function, respectively.
In order to harmonize   notations with \ref{IPA} and reference \cite{FZWard03}, 
we have chosen the units of mass in equations \eqref{sigth} and  \eqref{Sth}  so that 
$m=1$.

Though the integrals \eqref{sigth} and  \eqref{Sth} both increase linearly  at large $r$, their 
difference  vanishes exponentially at $r\to\infty$. The 
remaining radial integration in \eqref{d2pk} leads to the explicit representation for the coefficient 
$a_2$ in expansion \eqref{mqS},
\begin{equation}\label{a2int}
a_2=\frac{\pi}{2 \bar{s}^2}\int_0^\infty dr\, r[\mathcal{W}(r)-\mathcal{U}(r)].
\end{equation}
Numerical evaluation of this integral  yields
\begin{equation}\label{Aa2}
a_2=   0.0710108\ldots, 
\end{equation}
in agreement with \eqref{aq}.

The described calculation procedure is based both on the summation of the infinite form factor
series \eqref{d22}, and on the explicit representations for the matrix elements of the 
product of two spin operators between the one-fermion states, derived by 
Fonseca and Zamolodchikov in \cite{FZWard03}. 
Unfortunately, it is problematic  to extend  this approach to other integrable models, 
since it essentially exploits some rather specific features of the IFT, see the 'Discussion' Section in 
\cite{FZWard03}. On the other hand, a very good approximation for the constant $a_2$ can be obtained
by truncating the form factor series \eqref{d22} at its first term accounting  for
 the three-kink intermediate states, $n(q)=3$. We shall  describe this technique in some details here, and apply it in Section \ref{SOPFT} to estimate the leading quark-mass perturbative correction in the three-state PFT. 

The first term $\delta_{2,3} \,\langle \underline{p}|\mathcal{H}_R|\underline{k}\rangle$
in the form factor series \eqref{d22}, which describes contribution of the three-kink 
intermediate states has the following explicit form,
\begin{eqnarray}\label{H23pk}
&&\delta_{2,3} \,\langle \underline{p}|\mathcal{H}_R|\underline{k}\rangle=-\frac{1}{2}\frac{h^2}{3!}
\int_{-\infty}^\infty\frac{d q_1\,d q_2\,d q_3}{(2\pi)^3}\,\Delta(Q-p,R)\Delta(Q-k,R)\cdot
\\
&&\bigg\{\frac{ 1 }{\omega(q_1)+\omega(q_2)+\omega(q_3)- \omega(p)} +\frac{ 1 }{\omega(q_1)+\omega(q_2)+\omega(q_3)- \omega(k)} \bigg\}\, \nonumber
\cdot\\
&&\langle p|\sigma(0,0)|q_1, q_2,q_3\rangle
\langle q_3, q_2,q_1|\sigma(0,0)|k\rangle,\nonumber
\end{eqnarray}
where $Q=q_1+q_2+q_3$, and
\begin{equation}
\Delta(z,R)=\frac{2\sin(z R/2)}{z}.
\end{equation}
Note, that the two $R$-dependent factors in the integrand in \eqref{H23pk} give rise
to the momentum conservation law in the large-$R$ limit,
\begin{equation}\label{mcons}
\lim_{R\to\infty}\Delta(Q-p,R)\,\Delta(Q-k,R)=4\pi^2 \,\delta(Q-p)\,\delta(k-p).
\end{equation}

The right-hand side of \eqref{H23pk} is a well-defined generalized function 
for all finite $R$ under the conditions
\eqref{pkm}.  Exploiting the Wick expansion, the product of two matrix elements in the third line of \eqref{H23pk} can be represented as  the sum of nine terms. 
Taking into account the symmetry of the integrand in \eqref{H23pk} with respect to 
permutations of momenta  of  three virtual kinks, one can leave only two 
terms in this expansion multiplied by appropriate  combinatoric factors. As the result, 
 the substitution
\begin{eqnarray}\label{subs}
&&\langle p|\sigma(0,0)|q_1, q_2,q_3\rangle
\langle q_3, q_2,q_1|\sigma(0,0)|k\rangle \leadsto\\\nonumber
&&6\, \bar{\sigma}^{-2}\langle 0|\sigma(0,0)|q_2,q_3\rangle
\langle q_2,q_1|\sigma(0,0)|0\rangle\langle p|\sigma(0,0)|q_1\rangle
 \langle q_3|\sigma(0,0)|k\rangle+\\
&&3\,\bar{\sigma}^{-2}\langle 0|\sigma(0,0)|q_2,q_3\rangle
 \langle q_3,q_2|\sigma(0,0)|0\rangle
 \langle p|\sigma(0,0)|q_1\rangle  \langle q_1|\sigma(0,0)|k\rangle\nonumber
\end{eqnarray}
in the integrand in \eqref{H23pk} leaves the   integral unchanged.

One cannot proceed directly  to the limit $R\to\infty$ 
in equation \eqref{H23pk} exploiting equality
\eqref{mcons}. The problem comes from the product of two  kinematic singularities 
in the form factors in the 
right-hand side of \eqref{subs},
\begin{eqnarray}\label{ffct}
&& \langle p|\sigma(0,0)|q_1\rangle  \langle q_1|\sigma(0,0)|k\rangle
= \\\nonumber
&&- \bar{\sigma}^2
\, \frac{\omega(p)+\omega(q_1)}{\sqrt{\omega(p)\omega(q_1)}}\,
\frac{\omega(q_1)+\omega(k)}{\sqrt{\omega(q_1)\omega(k)}}\,
\mathcal{P}\frac{1}{p-q_1}\,\mathcal{P}\frac{1}{q_1-k}=\\\nonumber
&&4\pi^2  \bar{\sigma}^2 \,\delta(p-k)\delta(p-q_1)
+\left[\langle p|\sigma(0,0)|q_1\rangle  \langle q_1|\sigma(0,0)|k\rangle\right]_{reg},
\nonumber
\end{eqnarray}
where
\begin{eqnarray}\label{regpr}
\left[\langle p|\sigma(0,0)|q_1\rangle  \langle q_1|\sigma(0,0)|k\rangle\right]_{reg}=
-\frac{\bar{\sigma}^2}{2}
\frac{\omega(p)+\omega(q_1)}{\sqrt{\omega(p)\omega(q_1)}}\,
\frac{\omega(q_1)+\omega(k)}{\sqrt{\omega(q_1)\omega(k)}}\cdot\\\nonumber
\left[\frac{1}{(p-q_1-i0)(k-q_1-i0)}+\frac{1}{(p-q_1+i0)(k-q_1+i0)}
\right].
\end{eqnarray}
Multiplication of the first term in the right-hand side of \eqref{ffct} 
representing the direct propagation part  by the right-hand side 
 of equations  \eqref{mcons} leads to
the familiar meaningful factor $[\delta(p-k)]^2$.  This is not surprising,
since the vacuum energy 
$E_{vac}(m,h,R)\sim R$ contributing to 
$\langle \underline{p}|\mathcal{H}_R|\underline{k}\rangle$
diverges in the limit $R\to\infty$.

One can easily see, that the direct propagation part of the form factors \eqref{ffct}, 
upon substitution into \eqref{subs} and \eqref{H23pk}, gives rise
 to the term 
\begin{equation}\label{sing}
 2\pi\,\delta(p-k) \,\delta_{2,2}\,E(m,h,R),
\end{equation}
 where $\delta_{2,2}\,E(m,h,R)$
was defined in \eqref{a22L}. After subtraction of \eqref{sing}  from \eqref{H23pk}, we
 obtain a generalized function that has a well defined limit at $R\to\infty$. 
 According to \eqref{eps}, this limit must be identified with the 
 three-kink contribution to the second order correction to the kink dispersion law,
\begin{equation}\label{eps23}
2\pi \delta(p-k)\,\delta_{2,3} \,\epsilon(m,h,p)=\lim_{R\to\infty}
[\delta_{2,3} \,\langle \underline{p}|\mathcal{H}_R|\underline{k}\rangle-
2\pi\,\delta(p-k) \,\delta_{2,2}\,E(m,h,R)].
\end{equation}
After analytical continuation to all real $p$ and proceeding to the limit $p\to\infty$,
one obtains from \eqref{eps23} and \eqref{mq2}, the  corresponding correction to the squared kink mass
\begin{equation}\label{dm23}
\delta_{2,3}\, [m_q(m,h)]^2=m^2 \lambda^2 \, a_{2,3},
\end{equation}
where
\begin{equation}\label{A23}
a_{2,3}=\frac{1}{16\pi^2}\lim_{p\to\infty}[2\, \mathcal{I}_2(p)-\mathcal{I}_1(p)]
\end{equation}
is the three-kink contribution to the amplitude $a_2$. The explicit form of the integrals
$\mathcal{I}_{j}(p)$ reads as
\begin{eqnarray}\label{IJ}
\frac{\mathcal{I}_j(p)}{m^2}=\int_{-\infty}^\infty\frac{d q_1\,d q_2\,d q_3}{\omega(q_1)\omega(q_2)\omega(q_3)}
\frac{\delta(q_1+q_2+q_3-p)}{\omega(q_1)+\omega(q_2)+\omega(q_3)-\omega(p)}
\mathcal{J}_j(q_1,q_2,q_3),\\\label{II1}
\mathcal{J}_1(q_1,q_2,q_3)=\frac{[\omega(q_2)-\omega(q_3)]^2}{(q_2+q_3)^2}[\omega(p)+\omega(q_1)]^2\,\mathcal{P}\,\left(\frac{1}{p-q_1}\right)^2,\\\label{II2}
\mathcal{J}_2(q_1,q_2,q_3)=\frac{\omega(q_2)-\omega(q_1)}{q_2+q_1}
\frac{\omega(q_2)-\omega(q_3)}{q_2+q_3}
\left[\omega(p)+\omega(q_1)\right] \cdot\\
\left[\omega(p)+\omega(q_3)\right]\,\mathcal{P}\,\frac{1}{p-q_1}\, 
\mathcal{P}\,\frac{1}{p-q_3},\nonumber
\end{eqnarray}
where 
\begin{equation}\label{prv}
\mathcal{P}\,\left(\frac{1}{p-q}\right)^2=\frac{1}{2}\left[
\frac{1}{(p-k-i0)^2}+\frac{1}{(p-k+i0)^2}
\right].
\end{equation}
The constant \eqref{A23} was first numerically estimated by 
Fonseca and Zamolodchikov \cite{FonZam2003}, $a_{2,3}\approx 0.07$.
Its exact value 
\begin{equation}\label{aa23}
a_{2,3}=\frac{1}{16}+\frac{1}{12\pi^2}=0.0709434\ldots,
\end{equation}
which is remarkably close to the total amplitude $a_2$ [see (\ref{Aa2})],
was announced  later without derivation in \cite{Rut09}. To fill this gap, we 
present  the rather involved derivation of \eqref{aa23}  in \ref{Ca23}.

Finally, let us turn to  the  third-order term in the form factor expansion  \eqref{delta2}, 
and describe the main steps in  proof of equality \eqref{Hod} 
for $j=3$,  relegating  details  to \ref{LargeR}. 

We start from the form factor expansion  \eqref{delta3} 
and extract from it the 
direct propagation part, 
\begin{equation}\label{dppreg}
\delta_3 \langle \underline{p}|\mathcal{H}_R|\underline{k}\rangle=\delta_3 \langle \underline{p}|\mathcal{H}_R|\underline{k}\rangle_{dpp}+\delta_3 \langle \underline{p}|\mathcal{H}_R|\underline{k}\rangle_{reg}.
\end{equation}
After integration over ${\rm x}_1, {\rm x}_2, {\rm x}_3$ over the cube $(-R/2,R/2)^3$, we proceed in \eqref{dppreg}  to the 
limit $R\to\infty$ understood in the sense of generalized function. 
It turns out that only the direct propagation part of the matrix element \eqref{dppreg} contributes to this limit, 
giving rise to equality \eqref{Hod} at $j=3$, while the large-$R$ limit of its regular part 
 vanishes,
\begin{subequations}\label{qq}
\begin{align}
&\lim_{R\to\infty}\delta_3 \langle \underline{p}|\mathcal{H}_R|\underline{k}\rangle_{dpp}=-4\pi i \delta'(p-k)\delta_3\, \rho(m,h),
\label{RL}\\
&\lim_{R\to\infty}\delta_3 \langle \underline{p}|\mathcal{H}_R|\underline{k}\rangle_{reg}=0. \label{RL0}
\end{align}
\end{subequations}

\section{Form factors in the three-state PFT \label{FF}}
The form factors of physically relevant operators in the three-state PFT were found 
in 1988 by Kirillov and Smirnov in the preprint \cite{KS88} of the Kiev Institute for Theoretical Physics. In this section we briefly recall their results with emphasis on the 
form factors of the disorder spin operator in the paramagnetic phase. Exploiting the duality  
\cite{FRADKIN19801,Wu82,Bax} of the PFT, one can simply relate them to the form factors of the spin order operators in the ferromagnetic phase, which will by used in the next section. 

The set of nine operators operators $O_{ij}(x)$,  $i,j=0,1,2$ and their descendants were considered in \cite{KS88}.
The operators $O_{ij}$   transform in the following way under the action of the generator of the cyclic permutation 
$\Omega$ and charge conjugation $C$, 
\begin{equation}
\Omega^{-1} O_{ij} \Omega =\upsilon^i O_{ij}, \quad C^{-1}O_{ij}C=O_{\bar{i}\,\bar{j}},
\end{equation}
where $\upsilon=\exp(2\pi i/3)$, and $\bar{j}=3-j \mod 3$, $0\le \bar{j}\le 2$.
The operators $O_{ij}(x)$ were identified in \cite{KS88} 
as the main ones arising naturally  in the three-state PFT. 
In particular, the operators $O_{0j}$ with $j=1,2$ are proportional to the disorder spin operators \cite{FRADKIN19801}
$\mu$ and $\bar{\mu}$,
\begin{equation}\label{00j}
 O_{01}(x)=\frac{\mu(x)}{\langle \mu\rangle}, \quad  O_{02}(x)=\frac{\bar{\mu}(x)}{\langle \mu\rangle},
\end{equation}
where $\langle \mu\rangle=\!\!\!\phantom{x}_{par}\langle 0|\mu (0) |0\rangle_{par}=
\!\!\!\phantom{x}_{par}\langle 0|\bar{\mu}(0) |0\rangle_{par}$,
 and  $|0\rangle_{par}$ is the (non-degenerate) paramagnetic vacuum. The operators  $O_{0j}(x)$
 transform as scalars under  rotations. 
{ The operators $O_{j0}$,  ($j=1,2$) are proportional to the order spin operators $\sigma$ and $\bar{\sigma}$, respectively. 
The operators
$O_{jj}$,  ($j=1,2$)  correspond to parafermions $\psi_j$ (regularized $\sigma \mu$ and $\bar{\sigma} \bar{\mu}$), while $O_{j\bar{j}}$  $(j=1,2)$ 
are parafermions  $\bar{\psi}_j$ (regularized $\sigma \bar{\mu}$ and $\bar{\sigma }{\mu}$). Finally, the descendants
of the operator $O_{00}(x)$  correspond to the components of the energy-momentum density tensor and to other local conserved fields. The conformal limit of  these fields is described in \cite{ZF85}. }

We shall use notations \eqref{rapibas} for the 
3-state PFT rapidity 
basis states as well as  the normalization convention \eqref{norm3P} 
 in order to harmonize the notations with \cite{KS88}.
The form factors of the operator $O_{ij}(0)$  are defined as the matrix elements of the form
\begin{equation}\label{ff}
 f_{ij}(\beta_1,\ldots,\beta_n)_{\varepsilon_1,\ldots,\varepsilon_n}
 \equiv \!\!\! {\phantom{x}_{par} }\langle 0 | O_{ij}(0) | \beta_n,\ldots, \beta_1\rangle_{\varepsilon_n,\ldots,\varepsilon_1}.
\end{equation}
Due to their $\mathbb{Z}_3$-transformation properties, 
the form factors \eqref{ff} differ from zero only if $\sum_{k=1}^n \varepsilon_k = i  \mod 3 $.

The following axioms   \cite{smirnov1992form,KS88} are postulated for the form factors.
\begin{enumerate}
\item The symmetry property:
\begin{align}
f_{ij}(\beta_1,\ldots,\beta_l,\beta_{l+1},\ldots,\beta_n)_{\varepsilon_1\ldots,\varepsilon_l,\varepsilon_{l+1},\dots,\varepsilon_n}\, S_{\varepsilon_l,\varepsilon_{l+1}}(\beta_l-\beta_{l+1})=\\
f_{ij}(\beta_1,\ldots,\beta_{l+1},\beta_l,\ldots,\beta_n)_{\varepsilon_1\ldots,\varepsilon_{l+1},\varepsilon_l,\dots,\varepsilon_n}.\nonumber
\end{align}
\item The analytical continuation axiom:
\begin{align}
f_{ij}(\beta_1,\ldots,\beta_n+2\pi i)_{\varepsilon_1\ldots,\varepsilon_n}=\\
\upsilon^{-j\varepsilon_n}f_{ij}(\beta_n,\beta_1\ldots,\beta_{n-1})_{\varepsilon_n,\varepsilon_1,\ldots,\varepsilon_{n-1}}.\nonumber
\end{align}
\item The function $f_{ij}(\beta_1,\ldots,\beta_n)_{\varepsilon_1\ldots,\varepsilon_n}$ analytically
depends on the complex variables 
$\beta_n$ and has only simple poles in the strip $0\le \Im \beta_n \le \pi$ located at the 
points $\beta_n=\beta_k+\frac{2\pi i}{3}$, and $\beta_n=\beta_k+\pi i$. The residues at these points are:
\begin{align}\label{Bpole}
(2\pi)^{1/2}\,3^{-1/4}
{\rm Res}_{\beta_n=\beta_k+2\pi i/3} \,f_{ij}(\beta_1,\ldots,\beta_n)_{\varepsilon_1\ldots,\varepsilon_n}=\\
\delta_{\varepsilon_n,\varepsilon_k} \,f_{ij}(\beta_1,\ldots,\beta_k+\frac{\pi i}{3},\ldots,\beta_{n-1})_{\varepsilon_1\ldots,-\varepsilon_k,\ldots,\varepsilon_{n-1}}\cdot \nonumber\\
\prod_{l>k}^{n-1}S_{\varepsilon_n,\varepsilon_l}\left(\beta_k-\beta_l+\frac{2\pi i}{3}\right),\nonumber\\
2\pi i \,{\rm Res}_{\beta_n=\beta_k+\pi i } \,f_{ij}(\beta_1,\ldots,\beta_n)_{\varepsilon_1\ldots,\varepsilon_n}=\\
\delta_{\varepsilon_n,-\varepsilon_k} \,f_{ij}(\beta_1,\ldots,\hat{\beta}_k,\ldots,\beta_{n-1})_{\varepsilon_1\ldots,,\hat{\varepsilon}_k,\ldots,\varepsilon_{n-1}}\cdot \nonumber\\
\left\{
\prod_{l>k} S_{\varepsilon_l,\varepsilon_k}(\beta_l-\beta_k)-\upsilon^{\varepsilon_k j}
\prod_{l<k} S_{\varepsilon_k\varepsilon_l}(\beta_k-\beta_l)
\right\}. \nonumber
\end{align}
\end{enumerate}

The calculation of the  form factors 
$f_{ij}(\beta_1,\ldots,\beta_n)_{\varepsilon_1\ldots,\varepsilon_n}$ determined by the above axioms
was performed by Kirillov and Smirnov in  \cite{KS88}. Here we describe their results
for the case $i=0$, and $j=1,2$. It follows from \eqref{Bpole}, that the form factor 
$f_{0j}(\beta_1,\ldots,\beta_n)_{\varepsilon_1,\ldots,\varepsilon_n}$ can be expressed in terms of
$f_{0j}(\beta_1,\ldots,\beta_{3n})_{1,\ldots,1}$. The latter form factor will be denoted as
 $f_{0j}(\beta_1,\ldots,\beta_{3n})$. Its explicit representation reads as
\begin{align}\label{ff01}
f_{0j}(\beta_1,\ldots,\beta_{3n})=c^{-3n} g_{0j}(\beta_1,\ldots,\beta_{3n})\exp\left(-\frac{j}{3}\sum_{q=1}^{3n}\beta_q\right)\,\prod_{1\le l<k\le 3n}\zeta_{11}(\beta_l-\beta_k).
\end{align}
Here 
\begin{equation}
c=-i \sqrt{2\pi}\,3^{-1/12} \,\exp \left[
\frac{\psi^{(1)}(1/3)-\psi^{(1)}(2/3)}{12\sqrt{3}\,\pi}
\right]=-i\cdot 2.5474074563745797...  , 
\end{equation}
where $\psi^{(1)}(z)=\frac{d^2}{dz^2}\ln \Gamma(z)$ is the polygamma  function.
The function $\zeta_{11}(\beta)$ is defined  by the integral representation
\begin{align}\label{zeta11}
\zeta_{11}(\beta)=i \,2^{-2/3}\frac{\sinh ({\beta}/{2})}{\sinh[\frac{1}{2}(\beta-\frac{2\pi i}{3})]\sinh[\frac{1}{2}(\beta+\frac{2\pi i}{3})]} \cdot \\
\exp \left\{
2 \int_0^\infty dk \,\frac{\sin^2[\frac{1}{2}(\beta+ i \pi)k]+\frac{2}{3}\sinh^2(\pi k/3)}{k\, \sinh^2 (\pi k)} \sinh\frac{\pi k}{3}
\right\},\nonumber
\end{align}
which converges in the strip   $-8\pi/3<{\rm Im}\,\beta<2\pi/3$. This function can be analytically 
continuation into the whole complex $\beta$-plane, where it is meromorphic and
satisfies the equalities,
\begin{eqnarray}
\zeta_{11}(\beta)S_{11}(\beta)=\zeta_{11}(-\beta), \quad 
\zeta_{11}(\beta-2\pi i)=\zeta_{11}(-\beta),\\
\zeta_{11}\left(\beta-\frac{2\pi i}{3}\right)\,\zeta_{11}(\beta)\,
\zeta_{11}\left(\beta+\frac{2\pi i}{3}\right)=
\frac{1}{4\, \sinh\left(\frac{\beta}{2}-\frac{\pi i}{3}\right)\sinh\left(\frac{\beta}{2}+\frac{\pi i}{3}\right)}.\label{ze}
\end{eqnarray}  
The function  $\zeta_{11}(\beta)$ has a simple pole at $\beta=-{2\pi i/3}$ with the 
residue 
\begin{eqnarray}\label{resid}
{\rm Res}_{\beta=-{2\pi i/3}}\,\zeta_{11}(\beta)=
3^{1/6} i \exp \left[
\frac{\psi^{(1)}(1/3)-\psi^{(1)}(2/3)}{12\sqrt{3}\,\pi}
\right]=
-\frac{3^{1/4}c}{\sqrt{2\pi}}.
\end{eqnarray}
Note that in equations \eqref{zeta11} and \eqref{ze} we have corrected some misprints
which were present in \cite{KS88}. 

The functions $g_{0j}(\beta_1,\ldots,\beta_{3n})$ have the following representation,
\begin{equation}
g_{0j}(\beta_1,\ldots,\beta_{3n})=P_{0j,n}\left(
e^{\beta_1},\ldots ,e^{\beta_{3n}}
\right)\,\exp\left[
-(n-1)\sum_{q=1}^{3n}\beta_q
\right],
\end{equation}
where $P_{0j,n}(x_1,\ldots,x_{3n})$ is the uniform symmetric polynomial of the degree
$\deg(P_{0j,n})=3n^2-n\,\bar{j}$. 
The polynomial $P_{0j,n}(x_1,\ldots,x_{3n})$ can be represented as the determinant of the matrix $M_{0j,n}$ of the 
order $(2n-1) \times (2n-1)$, which has the matrix elements
\begin{equation}
\left(
M_{0j,n}
\right)_{pq}=\sigma_{3p-q-[(q-1+\bar{j})/2]}(x_1,\ldots,x_{3n}),
\end{equation}
where $[a]$ denotes the integer part of $a$, and 
$\sigma_k$ is the elementary symmetric polynomial of the variables 
$x_1,\ldots,x_{3n}$ of the degree $k$, and $\sigma_k=0$ for $k<0$, and for $k>3n$.

The first polynomials $P_{0j,n}(x_1,\ldots,x_{3n})$ have the form,
\begin{subequations}
\begin{eqnarray}
&&P_{01,1}(x_1,x_2,x_{3})=\sigma_1\equiv x_1+x_2+x_{3},\\
&&P_{02,1}(x_1,x_2,x_{3})=\sigma_2\equiv x_1x_2+x_1x_3+x_2 x_{3},\\
&&P_{01,2}(x_1,\dots,x_{6})=\sigma_1\sigma_3\sigma_4-\sigma_4^2-\sigma_1^2\sigma_6,\\
&&P_{02,2}(x_1,\dots,x_{6})=\sigma_2\sigma_3\sigma_5-\sigma_5^2-\sigma_2^2\sigma_6.
\end{eqnarray}
\end{subequations}
Accordingly, the  form factors \eqref{ff01} with $n=0,1$ read as,
\begin{subequations}\label{ff3}
\begin{eqnarray}
&&f_{01}(\varnothing)=1,\\
&&f_{01}(\beta_1,\beta_2,\beta_{3})=c^{-3}
\big[
e^{(-\beta_1-\beta_2+2\beta_3)/3}+
e^{(-\beta_1-\beta_3+2\beta_2)/3}+\\
&&e^{(-\beta_2-\beta_3+2\beta_1)/3}
\big]\prod_{1\le l<k \le 3}\zeta_{11}(\beta_l-\beta_k), \nonumber\\
&&f_{02}(\beta_1,\beta_2,\beta_{3})=c^{-3}
\big[
e^{(\beta_1+\beta_2-2\beta_3)/3}+
e^{(\beta_1+\beta_3-2\beta_2)/3}+\\
&&e^{(\beta_2+\beta_3-2\beta_1)/3}
\big]\prod_{1\le l<k \le 3}\zeta_{11}(\beta_l-\beta_k).\nonumber
\end{eqnarray}
\end{subequations}

The matrix elements of general form can be constructed from the form factor 
by means of the crossing relations \cite{smirnov1992form,KS88}. In particular,
\begin{eqnarray}\label{crossing}
\phantom{x}_{-1}
 \langle \beta |\mu(0)|\beta_2,\beta_1\rangle_{11}=\!\!\!
   {\phantom{x}_{par} }\langle 0| \mu(0)|\beta_2,\beta_1,\beta-i\pi\rangle_{111}=\\\nonumber
  \langle\mu\rangle f_{01}(\beta-i\pi,\beta_1,\beta_2),\\\label{crossing2}
  \phantom{x}_{-1} \langle \beta |\bar{\mu}(0)|\beta_2,\beta_1\rangle_{11}=
  \!\!\! {\phantom{x}_{par} }\langle 0| \bar{\mu}(0)|\beta_2,\beta_1,\beta-i\pi\rangle_{111}=\\
 \langle\mu\rangle f_{02}(\beta-i\pi,\beta_1,\beta_2).\nonumber
\end{eqnarray}

The above matrix elements of the disorder  operators relate to the paramagnetic phase. 
 { Let us connect them with the matrix elements of the order spin operators
 in the ferromagnetic phase. This can be easily done 
by means of the duality relations 
\begin{subequations}\label{dm}
\begin{eqnarray}\label{dm1}
\mu(x) \mathcal{D}=\mathcal{D}\sigma(x),\\
 \bar{\mu}(x) \mathcal{D}=\mathcal{D}\bar{\sigma}(x),
\end{eqnarray}
\end{subequations}
which connect the order  and disorder spin operators.  It is implied in \eqref{dm}  that the order spin operators 
$\sigma(x), \bar{\sigma}(x)$ act in the subspace $\mathcal{L}_0 $ of the ferromagnetic space
$\mathcal{L}_{fer}$, while the disorder spin operators
$\mu(x), \bar{\mu}(x)$ act in the subspace $\mathcal{L}_{sym}$ of the paramagnetic space $\mathcal{L}_{par}$.
All these vector spaces were described in Section \ref{PFTsec}.
Since $|\beta_2,\beta_1,\beta-i\pi\rangle_{111}\in \mathcal{L}_{sym}$, we can represent this vector as 
\[
|\beta_2,\beta_1,\beta-i\pi\rangle_{111}=\mathcal{D}\,
 |K_{02}(\beta_2)K_{21}(\beta_1)K_{10}(\beta-i\pi)\rangle.
\]
After substitution of this equality into \eqref{crossing} and straightforward manipulations exploiting
\eqref{dm1} and unitarity of the mapping $\mathcal{D}$, one obtains
\begin{equation*}
\phantom{x}_{par}\langle 0| \mu(0)||\beta_2,\beta_1,\beta-i\pi\rangle_{111}=\!\!\phantom{x}_{0}\langle 0|\sigma(0)|
K_{02}(\beta_2)K_{21}(\beta_1)K_{10}(\beta-i\pi)\rangle.
\end{equation*}
Application of the crossing relation\begin{footnote} {The crossing relations in the ferromagnetic PFT was discussed by 
Delfino and Cardy in the Appendix A of reference \cite{DelCard98}.}
\end{footnote}
to the right-hand side yields
\begin{equation*}
\phantom{x}_{0}\langle 0|\sigma(0)|
K_{02}(\beta_2)K_{21}(\beta_1)K_{10}(\beta-i\pi)\rangle=
\langle K_{10}(\beta)|\sigma(0)|
K_{02}(\beta_2)K_{21}(\beta_1)\rangle.
\end{equation*}
The right-hand side can be further transformed to the form
\[
\langle K_{10}(\beta)|\sigma(0)|
K_{02}(\beta_2)K_{21}(\beta_1)\rangle=\upsilon\,\langle K_{02}(\beta)|\sigma(0)|
K_{21}(\beta_2)K_{10}(\beta_1)\rangle,
\]
exploiting the transformation rule  $\sigma(0)=\upsilon\,\tilde{\Omega}\,\sigma(0)\,\tilde{\Omega}^{-1}$, and 
\eqref{OmK}.
Thus, we obtain finally from the above analysis,
\begin{equation}\label{simu}
\phantom{x}_{-1}
 \langle \beta |\mu(0)|\beta_2,\beta_1\rangle_{11} \big|_{par}=\upsilon\,\langle K_{02}(\beta) 
 |\sigma(0)| K_{21}(\beta_{2}) K_{10}(\beta_{1}) \rangle \big|_{fer}.
\end{equation}
Similarly, one can connect the matrix elements of the operators $\bar{\mu}(0)$ and $\bar{\sigma}(0)$,
\begin{equation}\label{simut}
\phantom{x}_{-1}
 \langle \beta |\bar{\mu}(0)|\beta_2,\beta_1\rangle_{11}\big|_{par}= \upsilon^{-1}\,\langle K_{02}(\beta) 
 |\bar{\sigma}(0)| K_{21}(\beta_{2}) K_{10}(\beta_{1}) \rangle \big|_{fer}.
\end{equation}
Combining  \eqref{simu}, \eqref{simut} }
 with \eqref{crossing}, \eqref{ff3} we find the three-kink matrix element
of the order operator
$
\sigma_3(0)=(\sigma(0)+\bar{\sigma}(0))/{3}
$
in the ferromagnetic phase, which will be used in the next Section,
\begin{eqnarray}\label{MEsig}
\langle K_{{0}2}(\beta) 
 |\sigma_3(0)| K_{21}(\beta_{2}) K_{1{0}}(\beta_{1}) \rangle \big|_{fer}=\\\nonumber
 \frac{\langle\mu\rangle}{3  c^3}\,\zeta_{11}(\beta-\beta_1-i \pi)\, \zeta_{11}(\beta-\beta_2-i \pi)\, \zeta_{11}(\beta_1-\beta_2)\\\nonumber
 \times\Bigg\{\left(
 e^{\beta_1} +e^{\beta_2}- e^{\beta}
 \right)\exp\left[-\frac{\beta_1+\beta_2+\beta_3+\pi i}{3}\right]+\\
\left(
 e^{-\beta_1} +e^{-\beta_2}- e^{-\beta}
 \right)\exp\left[\frac{\beta_1+\beta_2+\beta_3+\pi i}{3}\right]
 \Bigg\}.\nonumber
\end{eqnarray}
Note that the function      $\zeta_{11}(\beta)$ defined by equation \eqref{zeta11}
admits the following explicit representation in terms of the dilogarithm function 
${\rm Li}_2(z)=\sum_{n=1}^\infty \frac{z^n}{n^2}$,
\begin{eqnarray}\label{z11dL}
\zeta_{11}(\beta-i\pi)=-e^{-\beta/3}\left(1+e^{-\beta}\right)\left(1-e^{-\beta}+e^{-2\beta}\right)^{-5/6}\\
\times \left(
\frac{e^{\beta}-e^{i\pi/3}}{e^{\beta}-e^{-i\pi/3}}
\right)^{\frac{i\beta}{2\pi}}
\exp\left\{
\frac{i}{2\pi}\left[
{\rm Li}_2\left(e^{-\beta-\frac{i\pi}{3}}\right)-{\rm Li}_2\left(e^{-\beta+\frac{i\pi}{3}}\right)
\right]
\right\}.\nonumber
\end{eqnarray} 
The function in the right-hand side is even and real at real $\beta$. At  
${\rm Re}\,\beta\to+\infty$ 
it behaves as
\begin{eqnarray}\label{zetas}
\zeta_{11}(\beta-i\pi)=-e^{-\beta/3}\Bigg[1
+\frac{11\pi+3\sqrt{3}(1+\beta)}{6\pi}e^{-\beta}+\\
\frac{55\pi^2+27(1+\beta)^2+3\sqrt{3}\pi (25+28\beta)}{72\pi^2}e^{-2\beta}+
O\left(\beta^3e^{-3\beta}\right)\Bigg].\nonumber
\end{eqnarray}

To conclude this section, let us present a useful formula for the dilogarithm function
${\rm Li}_2(e^{i \pi p/q})$, with   $p< q$ for $p,q\in \mathbb{N}$:
\begin{equation}
{\rm Li}_2(e^{i \pi p/q})=\sum_{j=1}^q e^{i \pi j p/q} s(p,q,j),
\end{equation}
where
\begin{equation}
s(p,q,j)\equiv\sum_{l=0}^\infty\frac{e^{i \pi l p}}{(ql+j)^2}=
\begin{cases}
\frac{\psi^{(1)}(j/q)}{q^2}, & {\rm for \; even} \; p,\\
 \frac{\psi^{(1)}[j/(2q)]-\psi^{(1)}[(j+q)/(2q)]}{4q^2}, & {\rm for \; odd} \; p.
\end{cases}
\end{equation}
In particular,
\begin{equation}
{\rm Li}_2(e^{2 i \pi /3})=-\frac{\pi^2}{18}+i\,\frac{\psi^{(1)}(1/3)-\psi^{(1)}(2/3)}{6\sqrt{3}}.
\end{equation}
This equality has been used to derive  from \eqref{z11dL} the expression \eqref{resid} 
for the residue of the function $\zeta_{11}(\beta)$ at $\beta=-2\pi i/3$.
\section{Second-order quark mass correction in the ferromagnetic three-state PFT \label{SOPFT}}
In this section we estimate the second-order  radiative correction to the kink mass in the 
ferromagnetic 3-state PFT in the presence of a weak magnetic field $h>0$ coupled to the spin  component
 $\sigma_3$. Since very similar calculations for the case of the IFT were described in great  details  in Subsection \ref{1FS} and \ref{Ca23}, we can be brief. 

In the presence of the magnetic field, the Hamiltonian of the PFT associated with the action 
\eqref{AP} with $q=3$ takes the form
\begin{equation}
\mathcal{H}=\mathcal{H}_0-h\int_{-\infty}^\infty d{\rm x}\, \sigma_3({\rm x}),
\end{equation}
where the Hamiltonian $\mathcal{H}_0$ corresponds to the integrable ferromagnetic 3-state PFT 
at zero magnetic field. The kinks $K_{\mu\nu}(p)$ with the dispersion law 
$\omega(p,m)=\sqrt{p^2+m^2}$ are elementary excitations of the
model at $h=0$. For $h>0$, they  form mesonic and 
baryonic bound states at $h>0$ in the confinement regime. Nevertheless, one can determine  
the kink dispersion law $\epsilon(p,m,h)$ perturbatively in $h$, 
as  described in  Subsection  \ref{1FS}.
For the leading second order radiative correction $\delta_2 \,\epsilon(p,m,h)\sim h^2$ to the 
dispersion law of the kink $K_{2{0}}(p)$, one can write down the form factor expansion
\begin{eqnarray}\label{ffeP}
&&\delta_2 \,\epsilon(p,m,h)=\sum_{n=2}^\infty \delta_{2,n} \,\epsilon(p,m,h),\\\nonumber
&&\delta_{2,n} \,\epsilon(p,m,h)=-\frac{1}{n!}\frac{(2\pi)^2 h^2}{\omega(p)}\,
\sum_{\mu_1,\ldots,\mu_{n-1}= 0}^2
\int_{-\infty}^\infty \prod_{l=1}^n d\beta_l 
\frac{\delta(p_1+\ldots+p_n-p)}{\omega_1+\dots+ \omega_n-\omega}\\
&&\times|\langle K_{{0}2}(\beta) |\sigma_3(0)|K_{2, \mu_{n-1}}(\beta_{n})  K_{\mu_{n-1}, \mu_{n-2}}
(\beta_{n-1})   \dots     K_{\mu_{1},{0}}(\beta_{1})  \rangle|_{reg}^2,\label{deps2}
\end{eqnarray}
which is analogous to \eqref{d22}.  
Here $p_j= m \sinh \beta_j$, $\omega_j= m \cosh \beta_j$, 
$p=m\sinh\beta$, $\omega=  m\cosh\beta$. Of course, the same result holds for the kinks
 $K_{1{0}}(p)$, $K_{{0}2}(p)$ and $K_{{0}1}(p)$. 
 The  matrix elements  in the right-hand side of
\eqref{deps2} may contain kinematic singularities at $\beta_j=\beta$, which must be regularized 
as  done for the IFT  in   Subsection  \ref{1FS}. 
The second-order radiative
correction to the squared kink mass can be gained from $\delta_2 \,\epsilon(p,m,h)$ 
by taking the ultra-relativistic limit. Using  \eqref{mq2} gives
\begin{equation}
\delta_2 \,m_q(m,h)=2\lim_{p\to\infty} [\omega(p,m)\, \delta_2 \,\epsilon(p,m,h)].
\end{equation}

Let us truncate the  form factor expansion  \eqref{ffeP} at its first term with $n=2$,
\begin{eqnarray}\nonumber
\delta_{2,2} \,\epsilon(p,m,h)=-\frac{1}{2}\,\frac{(2\pi)^2 h^2}{\omega(p)}\, 
\int_{-\infty}^\infty  d\beta_1 d\beta_2 \,
\frac{\delta(p_1+p_2-p)}{\omega_1+ \omega_2-\omega}\\
\times|\langle K_{{0}2}(\beta) 
 |\sigma_3(0)| K_{21}(\beta_{2}) K_{1{0}}(\beta_{1}) \rangle|^2.\label{deps22}
\end{eqnarray}
The  matrix element in the right-hand side was calculated in the previous section,
see equation \eqref{MEsig}. Since  it is regular at all 
 real $\beta,\beta_1, \beta_2$, it does not  require regularization, in contrast to the subsequent terms 
in the expansion  \eqref{ffeP} with $n=3,4\ldots$.

\begin{figure}
\includegraphics{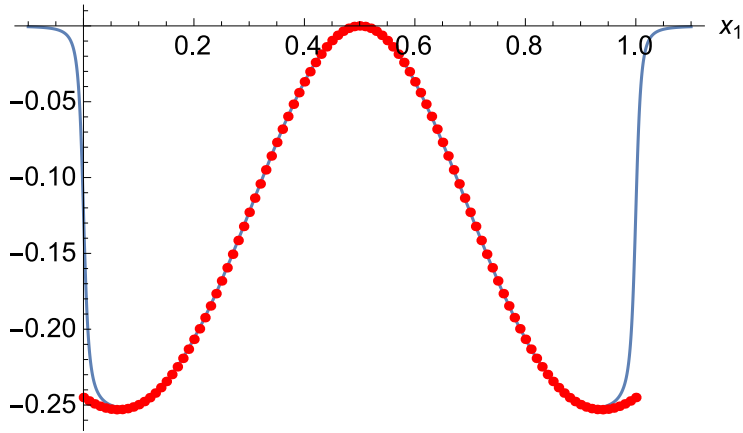}
\caption{Plot  of the  function  $\mathcal{M}(x_1,\mathfrak{p})$  defined by \eqref{a22A}, \eqref{a22B}  for $\mathfrak{p}=100$ (blue solid line), and of its ultra-relativistic limit  $\mathcal{M}(x_1,\infty)$  given 
by   \eqref{Minf}  (red circles). 
} \label{fig:M} 
\end{figure}

 The correction to the kink mass corresponding to \eqref{deps22} reads as
\begin{eqnarray}\label{dmP22}
\delta_{2,2} \,[m_q(m,h)]^2=
-{(2\pi  h)^2 }\, \lim_{\beta\to\infty}
\int_{-\infty}^\infty  d\beta_1 d\beta_2 \,
\frac{\delta(p_1+p_2-p)}{\omega_1+ \omega_2-\omega} \\
\times|\langle K_{{0}2}(\beta) 
 |\sigma_3(0)| K_{21}(\beta_{2}) K_{1{0}}(\beta_{1}) \rangle|^2.\nonumber
 \end{eqnarray}
Let us represent it in the form analogous to \eqref{mqS},
\begin{equation}\label{d22m2}
{\delta_{2,2} \,[m_q(m,h)]^2}=\lambda^2 a_{2,2} \,m^2,
\end{equation}
where $\lambda=f_0/m^2$ is the familiar dimensionless parameter proportional
to the magnetic field $h$, and
\begin{equation}
{ f_0= h [\!\!\!\phantom{x}_{0}\langle0| \sigma_3(0)|0\rangle_0
-\!\!\!\phantom{x}_{2}\langle0| \sigma_3(0)|0\rangle_2]= \frac{3}{2}h [\!\!\!\phantom{x}_{0}\langle0| \sigma_3(0)|0\rangle_{0}]
=h \langle 0|\mu(0)|0\rangle_{par}}
\end{equation}
 is the "bare" string tension
in the weak confinement regime. For the dimensionless  amplitude $a_{2,2}$, we obtain from 
\eqref{dmP22} and \eqref{MEsig},
\begin{eqnarray}\label{a22A}
&&a_{2,2}=-\frac{16\pi^2}{9 |c|^6} \lim_{\beta\to\infty}\int_{-\infty}^\infty  \,d\beta_1 d\beta_2\,
{\delta(\sinh \beta_1+\sinh \beta_2-\sinh \beta)}\\\nonumber
&&\times({\cosh \beta_1+\cosh \beta_2-\cosh \beta}) 
\left|\cosh\left(\frac{\beta_1+\beta_2+\beta+\pi i}{3}\right)\right|^2\\
&&\times|\zeta_{11}(\beta-\beta_1-i\pi)\zeta_{11}(\beta-\beta_2-i\pi)
\zeta_{11}(\beta_1-\beta_2)|^2.\nonumber
 \end{eqnarray}
After changing  the integration variables to $x_j=\sinh(\beta_j)/\sinh(\beta)$, $j=1,2$, 
and integrating over $x_2$  exploiting the $\delta$-function, one obtains
\begin{equation}\label{a22B}
a_{2,2}= \lim_{\beta\to\infty}\int_{-\infty}^\infty dx_1 \,\mathcal{M}(x_1,\sinh \beta).
\end{equation}
The function $\mathcal{M}(x_1,\mathfrak{p})$ is even with respect to the reflection
$x_1\to1-x_1$, and has the following asymptotic behavior at large ${\mathfrak{p}}\to\infty$, 
\begin{equation}
\mathcal{M}(x_1,\mathfrak{p})=\begin{cases}
\mathcal{M}(x_1,\infty)+O({\mathfrak{p}}^{-1}) , & {\rm for}\;0<x_1<1,\\
O(\mathfrak{p}^{-2}), & {\rm for}\;x_1<0, \; {\rm and \;for}\;x_1>1,
\end{cases}
\end{equation}
where
 \begin{eqnarray}\label{Minf}
\mathcal{M}(x_1,\infty)=-\frac{8\pi^2}{9 |c|^6}  \,\frac{1-x_1+x_1^2}{x_1^{4/3}(1-x_1)^{4/3}}\\
\times\left|\zeta_{11}(-\ln x_1-i\pi)\,\zeta_{11}[-\ln (1-x_1)-i\pi]
\,\zeta_{11}[\ln x_1-\ln (1-x_1)]\right|^2.\nonumber
 \end{eqnarray}
 Plots of $\mathcal{M}(x_1,\mathfrak{p})$ versus $x_1$ at $\mathfrak{p}=100$ and at $\mathfrak{p}=\infty$ are shown in Figure \ref{fig:M}.
 
 Thus, we arrive at the result
 \begin{equation}
  a_{2,2}=\int_{0}^1 dx_1 \,\mathcal{M}(x_1,\infty),
 \end{equation}
 with $\mathcal{M}(x_1,\infty)$ given by \eqref{Minf}. 
 We did not manage to evaluate the integral in the right-hand side 
 analytically, and instead computed it numerically 
 using \eqref{z11dL} and \eqref{zetas}. 
 The resulting number
 \begin{equation}\label{a22num}
 a_{2,2}=-\frac{4}{27}+\delta, \quad{\rm with}\; |\delta|<2\times 10^{-16}
 \end{equation}
is remarkably close to $-\frac{4}{27}$, which we assume to be the exact value of the amplitude $a_{2,2}$.
\section{Conclusions \label{Conc}}
In this paper we have  investigated the effect of the multi-quark (multi-kink) fluctuation 
on the universal characteristics of the IFT and 3-state PFT in the weak confinement 
regime, which is realized in these models in the low-temperature
phase in the presence of a weak magnetic field. For this purpose we  
refined the form factor perturbation technique which was adapted in  \cite{Rut09}  for  the 
confinement problem in the IFT.  
Due to  proper regularization of the merging kinematic singularities arising from the  products of spin-operator matrix elements,   the refined technique allowed us
to perform systematic high-order form factor perturbative  calculations in the weak  confinement 
regime.  After verifying the  efficiency of  the proposed method 
by  recovering several well-known results for the Ising model in the ferromagnetic phase in the 
scaling region, we have applied it to obtain  the following new results. 
\begin{itemize}
\item
The explicit  expression  \eqref{G32} for the contribution $\tilde{G}_{3,2}$ caused by two-quark 
fluctuations to the universal amplitude $\tilde{G}_{3}$, which characterizes the third derivative of the 
free energy of the scaling ferromagnetic Ising model  with respect to the magnetic field $h$ at $h=0$. 
\item
Proof of the announced earlier \cite{Rut09} exact result \eqref{aa23} for the amplitude $a_{2,3}$ describing the
contribution of three-quark fluctuations to the second order correction to the quark mass in the IFT
in the weak confinement regime. 
\item We showed that the third order $\sim h^3$ correction to the quark self-energy  
and to the quark mass vanishes in the  ferromagnetic IFT. 
This completes also calculations of the low-energy and semiclassical expansions for the meson masses 
$M_{n}(h,m)$ in the weak confinement regime to  third order in $h$. The final expansions for 
$M_{n}^2(h,m)$ to  third order in $h$ are described by the representations
given in \cite{Rut09}, since  only the terms (which are now shown to be zero)
proportional to  the third order quark mass corrections were missing there. 
\end{itemize}

In addition, a new representation \eqref{sigth}-\eqref{a2int} for the amplitude $a_2$ characterizing the second order 
radiative correction to the quark mass in the ferromagnetic IFT 
was obtained by performing the explicit integration over the polar angle in the
double-integral representation \eqref{a2B} for this amplitude obtained in \cite{FZWard03}.

Finally, exploiting the explicit expressions for the  form factors of the spin operators in the 3-state PFT  at zero magnetic field
obtained in \cite{KS88}, we have estimated the second-order radiative correction to the quark mass in the  ferromagnetic 
3-state PFT, which is induced by application of a weak magnetic field $h>0$.  To this end, we have truncated the
infinite form factor expansion for the second-order correction to the quark mass at its first term, which 
represents fluctuations with two virtual quarks in the intermediate state. 
Our result for the corresponding amplitude $a_{2,2}$ defined in \eqref{d22m2} is given in equations
 \eqref{Minf}-\eqref{a22num}, \eqref{z11dL}.

To conclude, let us mention two possible directions for further developments.  

Though the Bethe-Salpeter for the 
$q$-state PFT was obtained in paper \cite{Rut09}, it was not used there for the calculation of the meson mass spectrum. 
Instead, the latter  was  determined in  \cite{Rut09}  to the leading order in $h$ exploiting solely the zero-field scattering matrix known from 
\cite{CZ92}. The integral kernel in the Bethe-Salpeter for  the $q$-state PFT equation 
contains matrix elements of the spin operator $\sigma_q(0)$  between the two-quark states, that are not known for general $q$. In the case of $q=3$, however, such matrix elements
can be gained from the form factors found by Kirillov and Smirnov \cite{KS88}. This opens up the possibility to use
the Bethe-Salpeter equation for the 3-state PFT for analytical perturbative evaluation of the meson masses 
in subleading  orders in small $h$. On the other hand, one can also study the  magnetic field dependence of  
the  meson masses in the 3-state PFT at finite  magnetic fields by numerical solution of the Bethe-Salpeter equation. 
It was shown in \cite{FZ06} that  the Bethe-Salpeter equation reproduces surprisingly well
 the mesons masses in the IFT not only in the limit $h\to0$, but also at finite, and even at large
values of the magnetic field $h$. It would be interesting to check, whether this situation also takes 
place in the case of the $3$-state PFT. 

Recently, a dramatic effect of the kink confinement on the dynamics following a quantum quench 
was reported in \cite{RTak06,Kor16}  for  the IFT and for its discrete  analogue - the Ising chain in both
transverse and longitudinal magnetic fields. It was shown, in particular, that the masses of light mesons 
can be extracted from the spectral analysis of the post-quench time evolution of the one-point  functions. 
It would be interesting to extend these results to the 3-state PFT, in which 
both mesons and baryons are allowed. 
{\section*{Acknowledgements} 
\noindent 
I am grateful to A.~B.~Zamolodchikov for many important and stimulating discussions on the subject, 
and to  H.~W.~Diehl for interesting communications and numerous suggestions leading to improvement of the text.
I would like to thank F.~A.~Smirnov for sending me his preprint \cite{KS88}. 
In the initial stage, this work was supported by Deutsche Forschungsgemeinschaft (DFG) via Grant
Ru 1506/1.
\appendix
\section{Calculation of the integral \eqref{I2} \label{int2}}
Let us rewrite the double integral \eqref{I2} in the rapidity variables $\beta={\rm arcsinh}\,(q/m)$, 
$\beta'={\rm arcsinh}\,(q'/m)$,
\begin{eqnarray}\label{dint}
C_2=-\frac{1}{4}\int_{-\infty}^\infty d\beta \,
\frac{\sinh \beta }{\cosh^3 \beta} \int_{-\infty}^\infty d\beta' \,\frac{\sinh \beta' }{\cosh^3 \beta'} \cdot \\
\Bigg[
\coth^2\left(\frac{\beta-\beta'+i 0}{2}\right)+\coth^2\left(\frac{\beta-\beta'-i 0}{2}\right)
\Bigg], \nonumber
\end{eqnarray}
and consider the function 
\begin{equation}
u(\beta)=\int_{-\infty}^\infty d\beta' \,\frac{\sinh \beta' }{\cosh^3 \beta'}
\coth^2\left(\frac{\beta-\beta'}{2}\right)
\end{equation}
defined in the strip  $\Gamma=\left\{\beta\in \Gamma| \,0<{\rm Im}\,\beta<2\pi\right\}$. 
It is straightforward  to check its following properties.
\begin{enumerate}
\item The function $u(\beta)$ is analytic in the strip $\Gamma$ and vanishes there as $|\beta|\to\infty$.
\item $u( \pi i/2)= \pi i/4$, $u( 3\pi i/2)=- \pi i/4$, $u(\pi i)=0$.
\item The function $u(\beta)$ can be analytically continued to the whole complex $\beta$-plane, 
where it is meromorphic and obeys the quasi-periodicity relation
\begin{equation}
u(\beta+2\pi i)=u(\beta)-2\pi i \,v(\beta),
\end{equation}
with
\begin{equation}\label{vb}
v(\beta)=\frac{4\,[2-\cos(2 \beta)]}{\cosh^4 \beta}.
\end{equation}
\item The poles of $u(\beta)$ lie at $\frac{ \pi i}{2}+ i \pi n$ , with $n=-1,\pm 2,\pm 3,\dots$
\end{enumerate}
It is easy to prove that the above properties determine the function $u(\beta)$ uniquely, 
and to obtain its explicit form,
\begin{eqnarray}\label{ub}
u(\beta)=\frac{1}{4\cosh^4 \beta}\,\{
16(\beta-\pi i)[\cosh(2 \beta)-2]-\\23 \pi \sinh \beta -24 \sinh (2\beta)+\pi \sinh (3\beta)
\}.\nonumber
\end{eqnarray}
On the other hand, the double integral in equation \eqref{dint} defining the constant $C_2$ 
can be rewritten in terms of the functions $u(\beta)$, $v(\beta)$ as
\begin{equation}\label{C2}
C_2=-\frac{1}{4}\int_{-\infty}^\infty d\beta \,
\frac{\sinh \beta }{\cosh^3 \beta} 
\,[2\,u(\beta) +2\pi i \,v(\beta)
].
\end{equation}
After substitution of \eqref{vb} and \eqref{ub} into the right-hand side of \eqref{C2} and 
straightforward integration, one obtains  finally,
\begin{equation}
C_2=\frac{4}{3}+\frac{\pi^2}{8}.
\end{equation}
\section{Integration in the polar angle in \eqref{d2mq} \label{IPA}}
The subject of this Appendix is twofold. 
First, we prove that the representation \eqref{d2mq} for the second-order
radiative correction to the quark mass in the ferromagnetic IFT, which was derived in Section \ref{Is} in the frame  of the  modified form factor perturbative technique, is equivalent to the 
double-integral representation for the same quantity, which was obtained previously by Fonseca and Zamolodchikov, see  equations (5.6), (5.10) in \cite{FZWard03}.  Second, we perform analytical integration over the polar angle in the above-mentioned double-integral 
representation, and express the 
amplitude $a_2$ as a single integral in the radial variable $r$. 

In order to simplify our further notations, we shall 
 set throughout  Appendices B and C the units of mass, length and momentum so that 
\begin{equation}\label{m1}
m=1.
\end{equation}
The second order correction to the quark mass in the ordered phase  
is given by equation \eqref{d2mq}, 
which determines the dimensionless amplitude $a_2$ in expansion \eqref{mqS}, 
\begin{eqnarray}\nonumber
a_2=
-\frac{ 1}{2 \bar{s}^2}\lim_{\beta\to \infty}
\int_0^\infty d{\rm y} \int_{-\infty}^\infty d{\rm x}
\lim_{\beta'\to\beta} 
\langle \beta' | \sigma({\rm x},{\rm y})(1-{P}_1) \sigma(0,0)|\beta\rangle=
\\\label{a2}
-\frac{\pi }{2 \bar{s}^2}\lim_{\beta\to \infty}\int_0^\infty r \,dr\int_0^\pi \frac{d\theta}{\pi}
\left[G(r,\theta;\beta|\beta)-S_1(r,\theta;\beta|\beta)\right].
\end{eqnarray}
In the second line we have proceeded to the polar coordinates $r,\theta$ in the Euclidean half-plane, and used notations of \cite{FZWard03} for the matrix elements of the spin operators,
\begin{eqnarray}
G(r,\theta;\beta|\beta)=\lim_{\beta'\to\beta} \langle \beta' | \sigma({\rm x},{\rm y}) \sigma(0,0)|\beta\rangle,\\
S_1(r,\theta;\beta|\beta)=\lim_{\beta'\to\beta} \int_{-\infty}^\infty\frac{d \eta}{2\pi}
 \langle \beta' | \sigma({\rm x},{\rm y})|\eta\rangle\langle\eta| \sigma(0,0)|\beta\rangle,\label{S1}
\end{eqnarray}
where ${\rm x}=r\cos \theta$, 
${\rm y}=r\sin \theta$. 
 
Two further functions 
\begin{eqnarray}
&&S_{-1}(r, \theta;\beta|\beta)=\int_{-\infty}^\infty\frac{d \eta}{2\pi}\,
 \langle 0 | \sigma({\rm x},{\rm y})|\eta, \beta\rangle\langle\beta,\eta| \sigma(0,0)|0\rangle,\\
&& S_0(r, \theta;\beta|\beta)=S_{+1}(r,\theta;\beta|\beta)+S_{-1}(r, \theta;\beta|\beta)\label{S0}
\end{eqnarray}
will be  used in the sequel.
Functions $S_{\pm 1}(r, \theta;\beta|\beta)$,  describe the contributions of two different one-particle
reducible components in the matrix element 
$\langle \beta | \sigma({\rm x},{\rm y})\sigma(0,0)|\beta\rangle$, which were illustrated 
by two diagrams  in Fig. 3 in 
\cite{FZWard03}. Their explicit expressions read as
\begin{subequations}\label{Sa}
\begin{align}\label{S1a}
&S_{1}(r,\theta;\beta|\beta)=2 r  \bar{s}^2   \cosh(\beta+i\theta)+\\
&\bar{s}^2\int_{-\infty}^\infty \frac{d\eta}{2\pi}\,
\coth^2\left(\frac{\beta+i\theta-\eta}{2}\right)e^{i r [\sinh\eta-\sinh(\beta+i\theta)]},\nonumber\\
&S_{-1}(r, \theta;\beta|\beta)=\label{S2a}
 \bar{s}^2\int_{-\infty}^\infty \frac{d\eta}{2\pi}\,
\tanh^2\left(\frac{\beta+i\theta-\eta}{2}\right)e^{i r [\sinh\eta+\sinh(\beta+i\theta)]}.
\end{align}
\end{subequations}
The above representations hold for real $r, \beta, \theta$ lying in the region $r>0$, $-\infty<\beta<\infty$, and $0<\theta<\pi$, and can be extended to complex values of these variables
by analytical continuation. Note that the notation $S(r, \theta;\beta|\beta)$ was used  
 in \cite{FZWard03} for the function $S_0(r, \theta;\beta|\beta)$.

It follows from \eqref{S0} and  \eqref{Sa}  
that functions $S_j(r, \theta;\beta|\beta)$, with $j=0,\pm1$, depend in fact
on $r$  and the combination $\vartheta=\theta-i \beta$, being entire  functions of the complex variable $\vartheta$. These functions satisfy the following monodromy relations
\begin{equation}\label{mnS}
S_j(r, \pi;\beta|\beta)=S_{-j}(r, 0;\beta|\beta)-2 j r  \bar{s}^2 \sinh \beta, \quad {\rm for\;}j=0,\pm1.
\end{equation}
One can easily see from \eqref{S2a}, that the function $S_{-1}(r, \theta;\beta|\beta)$ vanishes 
in the limit $\beta\to\infty$ at fixed $r>0$ and $\theta\in [0,\pi]$. This allows one to replace 
the function $S_{1}(r, \theta;\beta|\beta)$ by $S_{0}(r, \theta;\beta|\beta)$ in the 
integrand in the second line in \eqref{a2},
\begin{eqnarray}
a_2=
\label{a2A}
-\frac{\pi }{2 \bar{s}^2}\lim_{\beta\to \infty}\int_0^\infty r \,dr\int_0^\pi \frac{d\theta}{\pi}
\left[G(r,\theta;\beta|\beta)-S_0(r,\theta;\beta|\beta)\right].
\end{eqnarray}

The  explicit expression 
for the function $G(r,\theta;\beta|\beta)$
 in terms of the pair correlation functions 
\begin{equation}\label{Iscor}
G(r)=\langle0| \sigma({\rm x},{\rm y}) \sigma(0,0)|0\rangle, \quad  \tilde{G}(r)=\langle0| \mu({\rm x},{\rm y}) \mu(0,0)|0\rangle,
\end{equation}
and associated auxiliary functions $\Psi_\pm(r,\vartheta)$ 
\begin{eqnarray}\nonumber
&&G(r,\theta;\beta|\beta)=i G(r)\left[\Psi_+(r,\vartheta)\,\partial_\vartheta \Psi_+(r,\vartheta)-
\Psi_-(r,\vartheta)\,\partial_\vartheta \Psi_-(r,\vartheta)\right]+\\\label{G2}
&& \tilde{G}(r) \Psi_+(r,\vartheta)\Psi_-(r,\vartheta),  \quad {\rm with}\; \vartheta=\theta-i\beta
\end{eqnarray}
were given in \cite{FZWard03}.  

The IFT correlation functions $G(r)$, $\tilde{G}(r)$ were found by Wu, McCoy, Tracy,  and Barouch in the classical paper \cite{McCoy76}. Properties of these and related functions 
$\Psi_\pm(r,\vartheta)$ are described  in much detail in \cite{FZWard03}.   Following this paper, we shall reproduce   some of these properties for  later use. 

The correlation functions \eqref{Iscor} admit the following representations
\begin{equation}\label{Is1}
G(r)=\exp[\chi(r)] \sinh[\varphi(r)], \quad  \tilde{G}(r)=\exp[\chi(r)] \cosh[\varphi(r)]
\end{equation}
in terms of the solutions of the ordinary Painlev\'e III differential equation,
\begin{eqnarray}\label{P3}
\varphi''(r)+\frac{1}{r}\varphi'(r)=\frac{1}{2} \sinh[2  \varphi(r)],\\
\chi''(r)+\frac{1}{r}\chi'(r)=\frac{1}{2}\left\{1-\cosh[2  \varphi(r)]\right\}.
\end{eqnarray}
The required solution is specified by its asymptotic behavior at $r\to0$,
\begin{eqnarray}
\varphi(r)=-\ln \frac{r}{2}-\ln(-\Omega)+O(r^4 \Omega^2),\\
\chi(r)=\frac{1}{2}\ln(4r) +\ln(-\Omega)+O(r^2),
\end{eqnarray}
where 
\begin{equation}
\Omega=\ln\left(
\frac{e^\gamma}{8} \,r
\right),
\end{equation}
and $\gamma$ is the Euler's constant. The solution $\varphi(r)$
decays at large $r\to\infty$ as
\begin{equation}
\varphi(r)=\frac{1}{\pi}\, K_0(r)+O(e^{-3r}).
\end{equation}

The auxiliary functions $\Psi_\pm(r,\vartheta)$ solve the system of partial differential
equations
\begin{subequations}\label{PsP}
\begin{eqnarray}
&& \partial_r \Psi_+=\frac{1}{4}\left(
 e^{\varphi+i\vartheta}- e^{-\varphi-i\vartheta}
 \right)  \Psi_-,\\
&& \partial_\vartheta \Psi_+=-\frac{i}{2}r \varphi'\,\Psi_++\frac{i}{4}\left(
 e^{\varphi+i\vartheta}+ e^{-\varphi-i\vartheta}
 \right) \, r \Psi_-,
\end{eqnarray}
\end{subequations}
and 
\begin{subequations}\label{PsM}
\begin{eqnarray}
&& \partial_r \Psi_-=-\frac{1}{4}\left(
 e^{-\varphi+i\vartheta}- e^{\varphi-i\vartheta}
 \right)  \Psi_+,\\
&& \partial_\vartheta \Psi_-=\frac{i}{2}r \varphi'\,\Psi_-  -\frac{i}{4}\left(
 e^{-\varphi+i\vartheta}+ e^{\varphi-i\vartheta}
 \right) \, r \Psi_+.
\end{eqnarray}
\end{subequations}
They are entire functions of the complex variable $\vartheta$ and  satisfy the monodromy properties
\begin{equation}\label{mnd}
\Psi_+(r,\vartheta+\pi)=i\,\Psi_+(r,\vartheta), \quad 
\Psi_-(r,\vartheta+\pi)=-i\,\Psi_-(r,\vartheta). 
\end{equation}
The following equality 
\begin{equation}\label{pspm}
\Psi_+(r,\vartheta)=\Psi_-(r,-\vartheta) 
\end{equation}
holds when both $r$ and $\vartheta$ are real. 
The overdetermined system of partial linear  differential equation \eqref{PsP}, \eqref{PsM} represents the Lax equations corresponding to  the nonlinear Painlev\'e III equation \eqref{P3}.

The function $G(r, \theta;\beta|\beta)$ depends in fact
on $r$  and the combination $\vartheta=\theta-i \beta$, being the entire $\pi$-periodical function of the complex variable $\vartheta$. The latter property leads to
the  equality
\begin{equation}\label{mnG}
G(r, 0;\beta|\beta)=G(r, \pi;\beta|\beta)
\end{equation}
for all real $\beta$ and $r$. 

Analyticity in $\vartheta$ and the monodromy properties \eqref{mnS}, 
 \eqref{mnG} guaranty 
that the integral 
\begin{equation}\label{thint}
\int_0^\pi \frac{d\theta}{\pi}
[G(r,\theta;\beta|\beta)-S_0(r,\theta;\beta|\beta)]
\end{equation}
in the right-hand side of \eqref{a2A} does not depend on $\beta$. This allows one to drop the 
$\lim_{\beta\to\infty}$ sign in equation \eqref{a2A}, yielding
\begin{eqnarray}\label{a2B}
a_2=
-\frac{\pi }{2 \bar{s}^2}\int_0^\infty r \,dr\int_0^\pi \frac{d\theta}{\pi}
\left[G(r,\theta;\beta|\beta)-S_0(r,\theta;\beta|\beta)\right],
\end{eqnarray}
with $\beta$-independent right-hand side. The integral representation \eqref{a2B} for the
amplitude $a_	2$  is equivalent to 
 equations (5.6), (5.10) obtained  for the 
 amplitude $a_q$ [see equation \eqref{aq1}] by Fonseca and Zamolodchikov in \cite{FZWard03}.

Now, let us proceed to the calculation of the integral \eqref{thint}. We start from the first term
\begin{equation}\label{UU}
\mathcal{U}(r)=\int_0^\pi \,\frac{d\theta}{\pi}
G(r,\theta;\beta|\beta). 
\end{equation}
Since the right-hand side does not depend on $\beta$, we shall put $\beta=0$ in it, and replace 
$\vartheta$ by $\theta$ in equations \eqref{G2}-\eqref{pspm}.

Let us introduce 
 three functions 
\begin{eqnarray}
f_1(r,\theta)&=&[\Psi_+(r,\theta)]^2\\\nonumber
f_2(r,\theta)&=&[\Psi_-(r,\theta)]^2,\\\nonumber
f_3(r,\theta)&=&\Psi_+(r,\theta)\Psi_-(r,\theta), 
\end{eqnarray}
which provide the 'spin-1' Lax representation 

\begin{eqnarray} \label{lax1}
-i\,\partial_\theta f_i(r,\theta)&=&\sum_{j=1}^3 f_j(r,\theta) U_{ji}(r,\theta) ,\\\nonumber
\partial_r f_i(r,\theta)&=&\sum_{j=1}^3 f_j(r,\theta) V_{ji}(r,\theta)
\end{eqnarray}
for the Painlev\'e III equation \eqref{P3}.  The matrices $U_{ji}(r,\theta)$ and $V_{ji}(r,\theta)$ are defined as 
\begin{eqnarray}
&&U(r,\theta)=\\
&&\begin{pmatrix}-r\,\varphi'(r)&0&- \frac{r}{4}\left[\frac{e^{\varphi(r)}}{v}+e^{-\varphi(r)} v\right]\\
0&r\varphi'(r)& \frac{r}{4}\left[e^{\varphi(r)}v+\frac{e^{-\varphi(r)}}{v}\right] \\
\frac{r}{2}\left[e^{\varphi(r)}v+\frac{e^{-\varphi(r)}}{v}\right]&
-\frac{r}{2}\left[\frac{e^{\varphi(r)}}{v}+e^{-\varphi(r)} v\right]&0
\end{pmatrix},\nonumber
\end{eqnarray}
\begin{eqnarray}
&&V(r,\theta)=\\
&&\begin{pmatrix}0 &0& \frac{1}{4}\left[\frac{e^{\varphi(r)}}{v}-e^{-\varphi(r)}v \right]\\
0&0& \frac{1}{4}\left[e^{\varphi(r)}v-\frac{e^{-\varphi(r)}}{v} \right]\\
\frac{1}{2}\left[{e^{\varphi(r)}}{v}-\frac{e^{-\varphi(r)}}{v}\right] &
\frac{1}{2}\left[\frac{e^{\varphi(r)}}{v}-{e^{-\varphi(r)}}{v}\right] &0
\end{pmatrix},\nonumber
\end{eqnarray}
where $v=e^{i\theta}$. The above 'spin-1' Lax equations can be easily deduced from the 
'spin-1/2'  Lax
equations (\ref{PsP}), \eqref{PsM}  for the functions $\Psi_\pm(r,\theta)$. 

Taking into account the 
symmetry properties \eqref{mnd}, \eqref{pspm},  the Fourier expansions for the 
functions $f_j(r,\theta)$ can be written in the form
\begin{eqnarray}\label{FE}
f_1(r,\theta)&=&\sum_{l=-\infty}^\infty  v^{2l+1} {\mathfrak a}_l(r)\\\nonumber
f_2(r,\theta)&=&\sum_{l=-\infty}^\infty  v^{-2l-1} {\mathfrak a}_l(r),\\\nonumber
f_3(r,\theta)&=&{\mathfrak b}_0(r)+\sum_{l=1}^\infty  \left(v^{2l}+ v^{-2l}\right)\,{\mathfrak b}_l(r).
\end{eqnarray}
Using equations (\ref{lax1}),  all Fourier coefficients  ${\mathfrak a}_l(r)$ and ${\mathfrak b}_l(r)$ can be  expressed
 recursively in terms of the coefficient ${\mathfrak b}_0(r)$
and its  derivative $ {\mathfrak b}_0'(r)$. 
The latter function solves the  second order linear differential equation \eqref{difb0}
which also follows from (\ref{lax1}). 

Asymptotical behavior of the function ${\mathfrak b}_0(r)$ at small and large $r$ can be gained from the 
known asymptotical behavior of the functions $\Psi_\pm(r,\theta)$ described in  \cite{FZWard03}. 
The result for small $r\to 0 $ reads as
\begin{equation}
{\mathfrak b}_0(r)=\frac{1}{\Omega}+r^4 g_4+r^8 g_8+ ...,
\end{equation}
where
\begin{eqnarray}
&&g_4=\frac{16 \Omega^3-8\Omega^2+1}{2^{11}\Omega^2},\\
&&g_8=\frac{8192 \Omega^6-12288 \Omega^5+7296 \Omega^4-1568\Omega^3-111\Omega+64}{2^{28}\Omega^3}.\nonumber
\end{eqnarray}
For the $r\to\infty$ asymptotics one finds,
\begin{equation}
{\mathfrak b}_0(r)=2\,{I}_0(r)+O(e^{-r}).
\end{equation}

Exploiting equations \eqref{lax1}, the function $G(r,\theta;0|0)$ determined by \eqref{G2} can be represented 
as a linear combination of functions $f_j(r,\theta)$,
\begin{eqnarray}
G(r,\theta;0|0)=\tilde{G}(r) f_3(r,\theta)+G(r)\Big[
f_3(r,\theta)r \varphi'(r)-\\
\frac{r}{4}\left(v\,  e^{-\varphi(r)}+v^{-1}e^{\varphi(r)}\right)f_1(r,\theta)
-\frac{r}{4}\left(v\,  e^{\varphi(r)}+v^{-1}e^{-\varphi(r)}\right)f_2(r,\theta)
\Big]. \nonumber
\end{eqnarray}
After substitution of the Fourier expansions \eqref{FE} in the right-hand side, the integration over 
the polar angle in \eqref{UU} becomes trivial. As the result, one represents the integral 
$\mathcal{U}(r)$
as a linear combination of the Fourier coefficients ${\mathfrak b}_0(r)$, ${\mathfrak a}_0(r)$, and ${\mathfrak a}_{-1}(r)$.
Expressing the  latter two coefficients in terms of ${\mathfrak b}_0(r)$  and  ${\mathfrak b}_0'(r)$, one arrives at the 
 result  given by equation \eqref{sigth}.

In order co complete the evaluation of the integral \eqref{thint}, it remains to calculate the 
second term,
\begin{equation}\label{WW}
\mathcal{W}(r) =\int_0^\pi \,\frac{d\theta}{\pi} \,S_0(r,\theta;\beta|\beta). 
\end{equation}
Since the right-hand side does not depend on $\beta$, we shall put $\beta=0$ in it.

Let us define an auxiliary function of the complex variable $\tilde{\beta}$,
\begin{eqnarray} \label{mf}
\mathfrak{f}(\tilde{\beta},r)=\int_{-\infty}^\infty \,\frac{d\eta}{2\pi}\,\exp[i r\,(\sinh \eta-\sinh\tilde{\beta})]\,
{\rm coth}\,\frac{\eta-\tilde{\beta}}{2},
\end{eqnarray}
where $0<{\rm Im}\, \tilde{\beta}<2\pi$, and the radius $r>0$ is fixed.  
The function $\mathfrak{f}(\tilde{\beta},r)$,  analytically continued to the whole complex 
$\tilde{\beta}$-plane,  satisfies there the quasi-periodicity relation
\begin{equation}\label{monf}
\mathfrak{f}(\tilde{\beta}+2\pi i,r)=\mathfrak{f}(\tilde{\beta},r)-2i.
\end{equation}
For the derivative $\partial_{\tilde{\beta}} \mathfrak{f}(\tilde{\beta},r)$, one can easily derive 
the following two representations from \eqref{mf},
\begin{equation} \label{dmf1}
\partial_{\tilde{\beta}} \mathfrak{f}(\tilde{\beta},r)=\frac{r}{\pi}
\left[i K_0(r) \sinh \tilde{\beta}-K_1(r)\right]\,\exp(-i r\,\sinh\tilde{\beta}),
\end{equation}
for all complex $\tilde{\beta}$, and 
\begin{eqnarray} \label{dmf2}
\partial_{\tilde{\beta}} \mathfrak{f}(\tilde{\beta},r)=-i r \,\mathfrak{f}(\tilde{\beta},r)\cosh \tilde{\beta}-\frac{K_0(r)\exp(-i r\,\sinh\tilde{\beta})}{2\pi}+\\
\frac{1}{2}
\int_{-\infty}^\infty \,\frac{d\eta}{2\pi}\,\exp[i r\,(\sinh \eta-\sinh\tilde{\beta})]\,
{\rm coth}^2\,\frac{\eta-\tilde{\beta}}{2}, \nonumber
\end{eqnarray}
for $0<{\rm Im}\, \tilde{\beta}<2\pi$. 

Comparison of  \eqref{dmf2} with \eqref{Sa} yields
\begin{subequations}\label{S12}
\begin{eqnarray}
 \frac{S_1(r,\theta;\beta|\beta)}{\bar{s}^{2}}=\Big\{2 r [1+i\mathfrak{f}(\tilde{\beta},r)]\cosh \tilde{\beta} +\\
\frac{K_0(r)e^{-i r\,\sinh\tilde{\beta}}}{\pi}+2\,\partial_{\tilde{\beta}}\mathfrak{f}(\tilde{\beta},r)
\Big\}\Big|_{\tilde{\beta}=\beta+i \theta}, \nonumber\\
 \frac{S_{-1}(r,\theta;\beta|\beta)}{\bar{s}^{2}}=\Big\{-2  i\,r\,\mathfrak{f}(\tilde{\beta}+i \pi,r)\,\cosh \tilde{\beta} +\\
\frac{K_0(r)e^{i r\,\sinh\tilde{\beta}}}{\pi}+2\,\partial_{\tilde{\beta}}\mathfrak{f}(\tilde{\beta}+i \pi,r)
\Big\}\Big|_{\tilde{\beta}=\beta+i \theta}.\nonumber
\end{eqnarray}
\end{subequations}
Upon adding these two  equalities and putting $\beta=0$ in the result, one finds,
\begin{eqnarray}
&& \frac{S_0(r,\theta;0|0)}{\bar{s}^{2}}=\Big\{2 r [1+i\mathfrak{f}(i \theta,r)-i\mathfrak{f}(i \theta+i \pi,r)]\cos {\theta} +\\
&&\frac{K_0(r)\left[e^{r\,\sin {\theta}}+e^{-r\,\sin {\theta}}\right]}{\pi}-2 i\,
\left[\partial_{ \theta}\mathfrak{f}(i \theta,r)+\partial_{ \theta}\mathfrak{f}(i \theta+i \pi,r)
\right]
\Big\}. \nonumber
\end{eqnarray}
Subsequent straightforward integration over  $\theta$ and use of
equalities \eqref{monf} and \eqref{dmf1} yields finally, 
\begin{eqnarray}\nonumber
&&\mathcal{W}(r)= \int_0^{\pi}\frac{d\theta}{\pi} S_0(r,\theta;0|0)=\\
&&\bar{s}^2 \int_0^{2\pi}\frac{d\theta}{\pi}
\Big\{2   i r\,\mathfrak{f}({i \theta},r) \cos {\theta} +
\frac{K_0(r)e^{r\,\sin \theta}}{\pi}-2 i\,
\partial_{\theta}\mathfrak{f}(i \theta,r)
\Big\}=\nonumber\\
&& \frac{2 \bar{s}^2}{\pi}
\bigg\{\left(1-2r^2\right) I_0(r) K_0(r)-
2 r\,K_1(r)\left[I_0(r)+r\, I_1(r)
\right]
\bigg\}.\label{WW3}
\end{eqnarray} 

\section{Calculation of $a_{2,3}$ \label{Ca23}}
In this Appendix we perform  the exact calculation of the amplitude $a_{2,3}$ given by 
equation \eqref{A23}, which characterize the 
three-kink contribution to the second-order radiative correction to the kink mass in the ferromagnetic IFT.  To this end, we evaluate  the integrals 
$\mathcal{I}_1(p)$ and  $\mathcal{I}_2(p)$ determined by equations \eqref{IJ}-\eqref{prv} in the limit $p\to\infty$, and 
show that
\begin{eqnarray}\label{limI1}
&&\lim_{p\to\infty}\mathcal{I}_1(p)=-\frac{4}{3}, \\
&&\lim_{p\to\infty}\mathcal{I}_2(p)=\frac{\pi^2}{2}.\label{limI2}
\end{eqnarray}
The momentum variables will be normalized throughout this 
Appendix to the "bare" kink mass according to the convention \eqref{m1}.

Proceeding to the calculation of the large-$p$ asymptotics of the integral $\mathcal{I}_1(p)$,
let us transform it to the variables  $x_j=q_j/p$, $j=1,2,3$, and 
expand the integrand in the right-hand side of \eqref{IJ} in small $1/p$ at fixed $x_j\ne0$. 
Since the energy denominator in it
\begin{eqnarray*}
{\omega(q_1)+\omega(q_2)+\omega(q_3)-\omega(p)}=
p(|x_1|+|x_2|+|x_3|-1)+\\
\frac{1}{2 p}\left(
\frac{1}{|x_1|}+\frac{1}{|x_2|}+\frac{1}{|x_3|}-1
\right)+O(p^{-3})
\end{eqnarray*}
becomes small $\sim p^{-1}$ on the part of the hyperplane defined by 
\begin{equation}\label{x123}
x_1+x_2+x_3=1, \quad {\rm with} \quad
x_{1,2,3}>0, 
\end{equation} 
let us assume  for a while  
that the leading contribution to the integral in the limit $p\to\infty$
comes from the region \eqref{x123}
\begin{footnote}
{
This assumption is not completely correct. As is shown below,  the vicinity of the point 
$x_1=1, \,x_2=0,\,x_3=0$ also gives a considerable contribution to the integral
${\mathcal{I}_{1}(p)}$ at  $p\to\infty$. 
}
\end{footnote}. Under this assumption, one obtains from \eqref{IJ}, 
\eqref{II1} at large $p$,
\begin{eqnarray}\label{I1a}
{\mathcal{I}_{1}(p)}=2 \int_0^\infty dx_1dx_2dx_3 \,
\frac{\delta(x_1+x_2+x_3-1)}{(x_2+x_3)(1-x_2)(1-x_3)}\cdot \\
\left(\frac{x_2-x_3}{x_2+x_3}\right)^2 
(1+x_1)^2\, \mathcal{P}\left(\frac{1}{1-x_1}\right)^2+O(p^{-1}).\nonumber
\end{eqnarray}
After trivial 
integration over $x_1$ and proceeding to the symmetric variables $u=x_2+x_3$, 
$v=x_2 x_3$, one obtains from \eqref{I1a},
\begin{eqnarray}\nonumber
{\mathcal{I}_{1}(p)}=4 \int_0^1 du \, \frac{(2-u)^2}{u^5}\int_0^{u^2/4}d v\, 
\frac{\left(u^2-4v\right)^{1/2}}{1-u+v}+O(p^{-1})=\\
4 \int_0^1 du \, \frac{(2-u)^2}{u^5}\label{I1b}
\left[
-2u+(u-2)\ln(1-u)
\right]+O(p^{-1}) .
\end{eqnarray}
The last integral diverges near its lower bound $u=0$.  
This divergence indicates
that the developed procedure cannot correctly 
describe the contribution of small momenta $|q_{2,3}|\ll p$ to the integral 
 $\mathcal{I}_1(p)$ defined by  \eqref{IJ},  \eqref{II1} in the limit $p\to\infty$.
 In order to regularize the integral $\int_0^1 du$ in the right-hand side of \eqref{I1b}, 
we split it into two terms as $\int_\epsilon^1 du+\int_0^\epsilon du$, where 
$\epsilon$ is an arbitrary small positive number. Thus,  $\mathcal{I}_1(p)$ 
becomes
 \begin{equation}\label{I1and2}
 \mathcal{I}_1(p)=\mathcal{I}_{1,>}(\epsilon)+ \mathcal{I}_{1,<}(\epsilon)+O(p^{-1}).
 \end{equation}
For the first term, we get
  \begin{equation}
 \mathcal{I}_{1,>}(\epsilon)=4 \int_\epsilon^1 du \, \frac{(2-u)^2}{u^5}\label{I1d}
\left[
-2u+(u-2)\ln(1-u)
\right]=\frac{8}{3\epsilon}-\frac{4}{3}+O(\epsilon).
 \end{equation}
We replace the second 
(diverging) integral $\mathcal{I}_{1,<}(\epsilon)$  in \eqref{I1and2} by the $p\to\infty$ limit of its 
converging finite-$p$ counterpart,
 \begin{equation}\label{lim12}
\lim_{p\to\infty} \mathcal{I}_1(p)=\mathcal{I}_{1,>}(\epsilon)+
\lim_{p\to\infty}  {\mathcal{I}_{1,<}(p,\epsilon p)},
 \end{equation}
where
\begin{eqnarray}\nonumber
&& {\mathcal{I}_{1,<}(p,q)}=\int_{-\infty}^\infty\frac{d q_1\,d q_2\,d q_3}{\omega(q_1)\omega(q_2)\omega(q_3)}
\frac{\delta(q_1+q_2+q_3-p)}{\omega(q_1)+\omega(q_2)+\omega(q_3)-\omega(p)}\cdot \\
&&\mathcal{J}_1(q_1,q_2,q_3) \,\eta(q- q_2-q_3).\label{I11}
\end{eqnarray}
Here  $\eta(z)$ stands for the unit-step function,
 \begin{equation}\label{step}
 \eta(z)=\begin{cases} 1, & {\rm for} \quad z>1,\\
 0, & {\rm for} \quad z\le 0,
 \end{cases}
 \end{equation}
 and $q=\epsilon p$ denotes  the cut-off momentum.

After integration over $q_1$ and proceeding to the limit $p\to \infty$ at a fixed positive $q$,
one obtains, 
\begin{equation} \label{I1pq}
{\mathcal{I}_{1,<}(p,q)}=4 p\, J_<(q)+O(p^{-1}),
\end{equation}
where
\begin{eqnarray} \label{Jless}
J_<(q)=\iint_{-\infty}^\infty\frac{dq_2 \,dq_3}{\omega(q_2)\,\omega(q_3)}\,
\frac{1}{\omega(q_2)-q_2+\omega(q_3)-q_3}\cdot\\
\left[\frac{
\omega(q_2)-\omega(q_3)}{q_2+q_3}\right]^2
\mathcal{P}\,\left(\frac{1}{q_2+q_3}\right)^2\,\eta(q- q_2-q_3).\nonumber 
\end{eqnarray}

First, let us show that the integral \eqref{Jless} vanishes, if the unit-step function in the 
integrand is dropped,
\begin{eqnarray}\label{Jin}
J\equiv \iint_{-\infty}^\infty\frac{dq_2 \,dq_3}{\omega(q_2)\,\omega(q_3)}\,
\frac{1}{\omega(q_2)-q_2+\omega(q_3)-q_3}\cdot\\
\left[\frac{
\omega(q_2)-\omega(q_3)}{q_2+q_3}\right]^2
\mathcal{P}\left(\frac{1}{q_2+q_3}\right)^2=0.\nonumber 
\end{eqnarray} 
Really, after a change of the integration variables to
\begin{equation} \label{xyvar}
x={q_2+\omega(q_2)+q_3+\omega(q_3)},\quad
y={[q_2+\omega(q_2)][q_3+\omega(q_3)]},
\end{equation}
we get
\begin{equation*}
J=8\int_0^\infty dy \, \mathcal{P}\, \frac{y^2}{(y-1)^2}\int_{2\sqrt{y}}^\infty
dx\, \frac{\sqrt{x^2-4 y}}{x^5}=
\frac{\pi}{16}\int_0^\infty dy \,\sqrt{y}\,\,\mathcal{P}\left( \frac{1}{y-1}\right)^2=0.
\end{equation*} 
Due to \eqref{Jin}, one concludes that 
\begin{equation} \label{Jgl}
J_<(q)=-J_>(q),
\end{equation}
where
\begin{eqnarray} \label{Jgr}
J_>(q)=  \iint_{-\infty}^\infty\frac{dq_2 \,dq_3}{\omega(q_2)\,\omega(q_3)}\,
\frac{1}{\omega(q_2)-q_2+\omega(q_3)-q_3}\cdot\\
\left[\frac{
\omega(q_2)-\omega(q_3)}{q_2+q_3}\right]^2
\,\frac{\eta(q_2+q_3-q)}{(q_2+q_3)^2},\nonumber 
\end{eqnarray} 
and it remains to calculate the large-$q$ asymptotics of the integral \eqref{Jgr}.
Transforming in this integral to the variables \eqref{xyvar}, one obtains
 \begin{equation} 
J_>(q)=  8\iint_0^\infty dx\,dy \, \frac{y^2}{(y-1)^2}
\frac{\sqrt{x^2-4 y}}{x^5}\,\eta(x^2-4 y)\eta[(y-1)x-2 q y ].
\end{equation}
After one more change of the integration variable $y=x^2 w$, we get 
\begin{eqnarray}\label{Jgrq}
J_>(q)=8 \int_0^{1/4} dw \, w^2	\sqrt{1-4w}\, X(w,q),
\end{eqnarray} 
where 
\begin{equation}
X(w,q)=\int_{x_0(w,q)}^\infty dx \,\frac{x^2}{(1-x^2 w)^2},
\end{equation}
and 
\[
x_0(w,q)={q+\sqrt{q^2+w^{-1}}}.
\]
Elementary integration in $x$ yields
\begin{equation*}
X(w,q)=
\frac{1+2 q \sqrt{w}\, {\rm arccoth}\left(q\, {w}^{1/2}+
\sqrt{1+q^2 w }\right)}{4 q w^2}.
\end{equation*}
Substitution of the large-$q$ asymptotics of this function
\begin{equation*}
X(w,q)=\frac{1}{2 q w^2}+O(q^{-3})
\end{equation*}
into \eqref{Jgrq} and subsequent integration over $w$ leads finally to the asymptotics
\begin{equation}
J_>(q)=\frac{2 }{3 q}+O(q^{-3})
\end{equation}
at  $q\gg 1$. Combining this result with 
 \eqref{Jgl} and \eqref{I1pq}, one obtains, 
\begin{equation} \label{I1pqA}
\lim_{p\to\infty}{\mathcal{I}_{1,<}(p,\epsilon p)}=-\frac{8 }{3 \epsilon}+O( \epsilon)
\end{equation}
at $\epsilon\ll1$. Adding \eqref{I1pqA} with \eqref{I1d}, we arrive 
at the result \eqref{limI1}.

Now let us proceed to the proof of equality \eqref{limI2}. Starting from the equations 
\eqref{IJ} and \eqref{II2}, which define the integral $\mathcal{I}_2(p)$, we first perform the integration
over $q_2$ by means of the $\delta$-function, then change the integration variables to
$x_j=q_j/p$, with j=1,3, and formally  proceed to the limit $p\to\infty$.  The result reads as
\begin{equation}
\lim_{p\to\infty }\mathcal{I}_2(p)=\int_{0}^1 dx_1 \int_0^{1-x_1} dx_3\, Y(x_1,x_3), 
\end{equation}
where
\[
Y(x_1,x_3)=2\,\frac{(1+x_1)(1+x_3)(1-2x_1-x_3)(1-x_1-2x_3)}{(1-x_1)^3(1-x_3)^3}.
\]
The double integral  in the right-hand side over the triangle $AOB$ 
shown in Figure \ref{trian} logarithmically diverges near the  edges $A$ and $B$ of the triangle. 
\begin{figure}[h]
\begin{pspicture}(11,9)
\psline[linewidth=1pt, linecolor=black]{->}(1,1)(1,8)
\psline[linewidth=1pt, linecolor=black]{->}(1,1)(8,1)
\psline[linewidth=2pt, linecolor=black]{-}(1,7)(7,1)
\psline[linewidth=2pt, linecolor=black]{-}(1,7)(1,1)
\psline[linewidth=2pt, linecolor=black]{-}(7,1)(1,1)
\psline[linewidth=1.2pt, linecolor=black]{-}(6,1)(6,2)
\psline[linewidth=1.2pt, linecolor=black]{-}(1,6)(2,6)
\psline[linewidth=1pt, linecolor=black]{-}(2,1)(2,.9)
\psline[linewidth=1pt, linecolor=black]{-}(6,1)(6,.9)
\psline[linewidth=1pt, linecolor=black]{-}(1,2)(.9,2)
\psline[linewidth=1pt, linecolor=black]{-}(1,6)(.9,6)
\rput[l]{00}(7.8,.6){$x_1$}
\rput[l]{00}(.4,7.9){$x_3$}
\rput[l]{00}(7.1,1.3){$A$}
\rput[l]{00}(6.8,.6){$1$}
\rput[l]{00}(.5,7){$1$}
\rput[l]{00}(1.3,7.1){$B$}
\rput[l]{00}(.6,.6){$O$}
\rput[l]{00}(.6,2.){$\epsilon$}
\rput[l]{00}(1.9,.6){$\epsilon$}
\rput[l]{00}(5.6,.6){$1-\epsilon$}
\rput[l]{00}(.1,6){$1-\epsilon$}
\pspolygon[fillcolor=gray,fillstyle=vlines](1,1)(1,6)(2,6)(6,2)(6,1)
\rput(3,3){\psframebox*[framearc=.3]{$\Gamma(\epsilon)$}}
\rput[l]{00}(7,1.7){$\Delta_A(\epsilon)$}
\psline[linewidth=.2pt]{->}(6.9,1.7)(6.3,1.3)
\rput[l]{00}(2.,6.7){$\Delta_B(\epsilon)$}
\psline[linewidth=.2pt]{->}(1.9,6.7)(1.3,6.3)
\end{pspicture}
\caption{Integration  regions in integrals \eqref{limY}, \eqref{DelA}.  \label{trian}}
\end{figure}
In order to regularize this integral, we divide the triangle AOB   
into the polygon $\Gamma(\epsilon)$, 
which is  dashed in Figure  \ref{trian},  and two small  rectangular triangles  $\Delta_{A,B}(\epsilon)$
 adjacent to the edges $A$ and $B$. 
The legs of  these small  triangles have the length $\epsilon$. 
The integral  over the 
polygon $\Gamma(\epsilon)$ approaches 
in the limit $\epsilon\to0$  a constant value,
\begin{equation}\label{limY}
\lim_{\epsilon\to0}\iint_{\Gamma(\epsilon)}dx_1 dx_3\, Y(x_1,x_3)=\frac{\pi^2}{2}.
\end{equation}

The similar integrals over small triangles adjacent to the points $A$ and $B$ are equal to one another, but  
formally diverge. To prove equation \eqref{limI2}, it remains to
 show that these integrals  vanish after regularization. 
To this end, let us consider the integral 
\begin{eqnarray}\nonumber
&&{\mathcal{I}_{2,A}(p,q)}=\int_{-\infty}^\infty\frac{d q_1\,d q_2\,d q_3}{\omega(q_1)\omega(q_2)\omega(q_3)}
\frac{\delta(q_1+q_2+q_3-p)}{\omega(q_1)+\omega(q_2)+\omega(q_3)-\omega(p)}\cdot \\
&&\mathcal{J}_2(q_1,q_2,q_3)\,\eta(q-q_2-q_3),\label{I2A}
\end{eqnarray} 
where $\mathcal{J}_2(q_1,q_2,q_3)$ is given by \eqref{II2}, and $q=\epsilon p$.
Clearly, this well-defined integral represents the  finite-$p$ 
regularized counterpart of the diverging integral 
\begin{equation}\label{DelA}
\int_{\Delta_A(\epsilon)} dx_1dx_3\, Y(x_1,x_3).
\end{equation}
After integration over $q_1$, one finds from \eqref{I2A} at fixed $q>0$ in the limit $p\to\infty$,
\begin{eqnarray}\nonumber
\lim_{p\to\infty}{\mathcal{I}_{2,A}(p,q)}=2\iint_{-\infty}^\infty\frac{d q_2\,d q_3}{\omega(q_2)\omega(q_3)}\,
\frac{1}{\omega(q_2)-q_2+\omega(q_3)-q_3}\cdot \\
\frac{\omega(q_2)-\omega(q_3)}{q_2+q_3}\,
\eta(q-q_2-q_3)\,\mathcal{P}\left(\frac{1}{q_2+q_3}\right)=0,
\end{eqnarray} 
 since the integrand in the right-hand side is odd with respect of the permutation $q_2\leftrightarrow q_3$.
 This completes the proof of  equations \eqref{limI2}.
\section{Proof of equations \eqref{qq} \label{LargeR}}
The form factor expansion \eqref{delta3} contains the product of three matrix elements of the spin operators
\begin{equation}\label{sig3}
\langle p |{\sigma}(0,0)|q\rangle\langle q|{\sigma}(0,0)|q'\rangle
\langle q'|{\sigma}(0,0)|k\rangle.
\end{equation}
Here $|q\rangle$ and $|q'\rangle$  stand  for the intermediate 
fermionic states with odd numbers of kinks, $n(q)$, and $n(q')$, respectively.
Note, that $n(q)+n(q')\ge 4$. By means of the Wick expansion, 
the product \eqref{sig3} can be brought into the sum of products 
of $n(q)+n(q')+1$ elementary form factors  \eqref{fIs}-\eqref{fIs3}.
 Some terms in this Wick expansion contain the products of three elementary
 form factors of the form 
\begin{equation}\label{sig3b}
 \langle p |{\sigma}(0,0)|q_i\rangle\langle q_i|\sigma(0,0)|q'_j\rangle
\langle q'_j|{\sigma}(0,0)|k\rangle,
\end{equation}
where $1\le i \le n(q)$, and $1\le i \le n(q')$. Extracting the direct propagation part from such a product,
one  can  represent it as  
\begin{eqnarray}   \label{sig3a}
\langle p |{\sigma}(0,0)|q_i\rangle\langle q_i|{\sigma}(0,0)|q'_j\rangle
\langle q'_j|{\sigma}(0,0)|k\rangle =\\\nonumber
4\pi^2 \langle p |{\sigma}(0,0)|k\rangle\, \delta(q_i-k) \, \delta(q_j'-k)+  \\\nonumber
 \langle p |{\sigma}(0,0)|q_i\rangle \big[  \langle q_i |{\sigma}(0,0) |q'_j\rangle
\langle q'_j|{\sigma}(0,0)|k\rangle  \big]_{reg}.
\end{eqnarray}
The regularized product of two elementary form factors standing in the last line 
was defined in \eqref{regpr}.  One can also extract the direct propagation part from the 
product \eqref{sig3b} in a different way
\begin{eqnarray}   \label{sig3c}
\langle p |{\sigma}(0,0)|q_i\rangle\langle q_i|{\sigma}(0,0)|q'_j\rangle
\langle q'_j|{\sigma}(0,0)|k\rangle =\\\nonumber
4\pi^2 \langle p |{\sigma}(0,0)|k\rangle\, \delta(q_i-p) \, \delta(q_j'-p)+  \\\nonumber
\big[  \langle p |{\sigma}(0,0)|q_i\rangle  \langle q_i |{\sigma}(0,0) |q'_j\rangle  \big]_{reg}\,
\langle q'_j|{\sigma}(0,0)|k\rangle.
\end{eqnarray}
Taking the arithmetic average of equations \eqref{sig3a} and \eqref{sig3c}, we get the right-hand side in the 
symmetrized form
\begin{eqnarray}   \label{sig3d}
\langle p |{\sigma}(0,0)|q_i\rangle\langle q_i|{\sigma}(0,0)|q'_j\rangle
\langle q'_j|{\sigma}(0,0)|k\rangle =\\\nonumber
\big[\langle p |{\sigma}(0,0)|q_i\rangle\langle q_i|{\sigma}(0,0)|q'_j\rangle
\langle q'_j|{\sigma}(0,0)|k\rangle\big]_{dpp}+  \\\nonumber
\big[\langle p |{\sigma}(0,0)|q_i\rangle\langle q_i|{\sigma}(0,0)|q'_j\rangle
\langle q'_j|{\sigma}(0,0)|k\rangle\big]_{reg},
\end{eqnarray}
where
\begin{eqnarray}  \label{dppk}
&&\big[\langle p |{\sigma}(0,0)|q_i\rangle\langle q_i|{\sigma}(0,0)|q'_j\rangle
\langle q'_j|{\sigma}(0,0)|k\rangle\big]_{dpp}=\\\nonumber
&&2\pi^2 \langle p |{\sigma}(0,0)|k\rangle\, 
[\delta(q_i-k) \, \delta(q_j'-k)+\delta(q_i-p) \, \delta(q_j'-p)],\\
&&\big[\langle p |{\sigma}(0,0)|q_i\rangle\langle q_i|{\sigma}(0,0)|q'_j\rangle
\langle q'_j|{\sigma}(0,0)|k\rangle\big]_{reg}=\\\nonumber
&&\frac{1}{2}\big[\langle q_i|{\sigma}(0,0)|q'_j\rangle
\langle q'_j|{\sigma}(0,0)|k\rangle\big]_{reg}\,\langle p |{\sigma}(0,0)|q_i\rangle
+\\\label{regpk}
&&\frac{1}{2}
\big[\langle p |{\sigma}(0,0)|q_i\rangle\langle q_i|{\sigma}(0,0)|q'_j\rangle\big]_{reg}\,
\langle q'_j|{\sigma}(0,0)|k\rangle\nonumber
\end{eqnarray}
Collecting all  terms in the Wick expansion, that contain the factors of the form \eqref{sig3b},
  and leaving in those only the symmetrized direct propagation parts \eqref{dppk},
one obtains the direct propagation part $\delta_3 \langle \underline{p}|\mathcal{H}_R|\underline{k}\rangle_{dpp}$
of the matrix element $\delta_3 \langle \underline{p}|\mathcal{H}_R|\underline{k}\rangle$. Denoting the rest
of the latter by 
 $\delta_3 \langle \underline{p}|\mathcal{H}_R|\underline{k}\rangle_{reg}$, we arrive to equation \eqref{dppreg}.
 
The matrix element 
$ \delta_3 \langle \underline{p}|\mathcal{H}_R|\underline{k}\rangle$ defines in a standard way a distribution, 
which acts on a 
 'good enough'  test function $\phi(p,k)\in {\mathscr{D}}_{\mathcal{S}}$ having a  compact support $\mathcal S$,
\begin{equation}\label{dsHdis}
\delta_3 {\mathcal H}_R[\phi]=\iint dp\, dk\, \delta_3 \langle \underline{p}|\mathcal{H}_R|\underline{k}\rangle\, \phi(p,k).
\end{equation}
To avoid the resonance poles, the support  of the test function will be taken inside the square, 
${\mathcal S}\subset  (-p_0,p_0)^2$ with $p_0=2^{3/2}m$.
Due to the symmetry relation \eqref{hh}, the test functions $\phi(p,k)$ can be chosen odd,
\begin{equation}\label{symphi}
\phi(p,k)=-\phi(k,p) \quad {\rm for}\quad \phi \in {\mathscr{D}}_{\mathcal{S}},
\end{equation}
without loss of generality.

Similarly to \eqref{dsHdis}, one can determine  the  action on  $\phi \in {\mathscr{D}}_{\mathcal{S}}$ 
of the distribution $\delta_3 {\mathcal H}_{R,{dpp}}$ associated 
with the direct propagation part  of the matrix element 
$ \delta_3 \langle \underline{p}|\mathcal{H}_R|\underline{k}\rangle$,
\begin{eqnarray}\label{acH}
&&\delta_3 {\mathcal H}_{R,dpp}[\phi]\equiv \iint dp\, dk\, \delta_3 \langle \underline{p}|\mathcal{H}_R|\underline{k}\rangle_{dpp}\, \phi(p,k)=\\
&&2{i}\,\iint dp \,dk \,\phi(p,k)\int\frac{dQ \,dQ' } {4\pi^2}\nonumber
D_1(p-k,Q, Q';R)\,
 {\mathcal{Y}}_{dpp}(p,k,Q,Q';m,h),
 \end{eqnarray}
 with
 \begin{equation} \label{DPQ}
 D_1(P,Q, Q';R)=\frac{8 \sin[(P-Q)R/2]\sin[(Q-Q')R/2]\sin[Q' R/2]}{(P-Q)(Q-Q') Q'}\,\mathcal{P}\,\frac{1}{P}.
\end{equation}
Here $Q$ and $Q'$ denote the total momenta of the intermediate kink states in the form factor expansion. 
The function $ {\mathcal{Y}}_{dpp}(p,k,Q,Q';m,h)$ in the right-hand side of \eqref{acH}, which is  
 analytic in its  momenta variables for $\{p,k\}\in \mathcal{S}$   and  all $Q,Q'$, has the 
 following symmetry properties
\begin{eqnarray}\label{symY}
 && {\mathcal{Y}}_{dpp}(p,k,Q,Q';m,h)={\mathcal{Y}}_{dpp}(k,p,Q',Q;m,h).\\
&& {\mathcal{Y}}_{dpp}(p,k,Q,Q';m,h)=  {\mathcal{Y}}_{dpp}(p,k,-Q,-Q';m,h),\nonumber
  \end{eqnarray}
 and reduces
 at $p=k$, and $Q=Q'=0$  
 to the third correction to the vacuum energy density \eqref{d3Rho}, 
 \begin{equation}\label{YY}
  {\mathcal{Y}}_{dpp}(k,k,0,0;m,h)=\delta_3 \rho(m,h).
 \end{equation}

It remains to proceed to the large-$R$ limit in equation \eqref{acH}. To this end, let us determine how 
 the distribution \eqref{DPQ} acts on the plane wave test function,
\begin{eqnarray}\label{JJ}
 J({\bf X};R)=\int_{-\infty}^\infty dP\, dQ \, dQ'\, D_1(P,Q, Q';R)\cdot\\
\exp \{i [P X_1+Q(X_2-X_1)+Q'(X_3-X_2)]\}\nonumber,
 \end{eqnarray}
 where $ {\bf X}=\{X_1,X_2,X_3\}$.
After the change of  integration variables 
\[
P=u_1+u_2+u_3, \quad Q=u_2+u_3,\quad Q'=u_3,
\]
the integral representation for the function $ J({\bf X};R)$  
takes a symmetric form, and can be easily calculated
\begin{eqnarray}\nonumber
 &&J({\bf X};R)=\int_{-\infty}^\infty du_1\, du_2 \, du_3\, {\mathcal P}\,\frac{1}{u_1+u_2+u_3}
\prod_{j=1}^3 \frac{2\,  e^{i u_j X_j}\,\sin(u_j  R/2) }{u_j}=\\\label{J3}
&& \frac{i}{2}  \int_{-\infty}^\infty d\lambda \, {\rm sign} \lambda \prod_{j=1}^3 
\int_{-\infty}^\infty du_j\,\frac{2\,  e^{i u_j (X_j-\lambda)}\,\sin(u_j  R/2) }{u_j}=\\\nonumber
&&4 \pi^3 i\,\eta(R-X_{max}+X_{min})\cdot\\
&&[ \max (X_{min},-R-X_{min})+\min (X_{max},R-X_{max})],\nonumber
 \end{eqnarray}
 where $\eta(z)$ is  the unit-step function \eqref{step}.
 In the  second line in \eqref{J3}, the integral representation  
 \[
{\mathcal P}\,\frac{1}{u}=\frac{i}{2}\int_{-\infty}^\infty  d\lambda\, e^{-i \lambda u}\, {\rm sign}\, \lambda
 \]
 has been used.
Proceeding to the limit $R\to\infty$ in \eqref{J3}, one obtains
 \begin{equation}\label{JR}
J({\bf X};\infty)=4 \pi^3 i\,(X_{max}+X_{min}).
 \end{equation}
This results indicates that the  distribution \eqref{DPQ} remains nonlocal in the limit $R\to\infty$. 
It turns out, however, that the large-$R$ limit of  \eqref{DPQ} determines the following  local distribution 
\begin{eqnarray}\label{loc}
\lim_{R\to\infty}\int_{-\infty}^\infty dP\, dQ \, dQ'\, D_1(P,Q, Q';R)\,
\Phi(P,Q,Q')=\\\nonumber
4\pi^2\left(2\,\partial_P+\partial_Q+\partial_{Q'}
\right)\Phi(P,Q,Q')\Big|_{P=Q=Q'=0},
 \end{eqnarray}
when it is restricted to the space of test functions, that satisfy the symmetry relation
\begin{equation}\label{symPh}
\Phi(P,Q,Q')=-\Phi(-P,-Q',-Q).
\end{equation}
To prove equality \eqref{loc}, it is sufficient to check that it holds for the  'anti-symmetrized plane-wave' test function
\begin{eqnarray*}
\exp \{i [P X_1+Q(X_2-X_1)+Q'(X_3-X_2)]\}-\\
\exp \{i [-P X_1-Q'(X_2-X_1)-Q(X_3-X_2)]\},
\end{eqnarray*}
which obeys  \eqref{symPh}. This can be easily done by application of \eqref{JR}.
 Combining \eqref{loc} with \eqref{symphi}-\eqref{YY}, we arrive at \eqref{RL}. 
 
The proof of equation  \eqref{RL0} is simpler. The regular part 
$ \delta_3 \langle \underline{p}|\mathcal{H}_R|\underline{k}\rangle_{reg}$ of the 
matrix element $ \delta_3 \langle \underline{p}|\mathcal{H}_R|\underline{k}\rangle$
was defined according to equation \eqref{dppreg} as
\begin{equation}\label{dppreg1}
\delta_3 \langle \underline{p}|\mathcal{H}_R|\underline{k}\rangle_{reg}=\delta_3 \langle \underline{p}|\mathcal{H}_R|\underline{k}\rangle-\delta_3 \langle \underline{p}|\mathcal{H}_R|\underline{k}\rangle_{dpp}.
\end{equation}
After integration over the variables ${\rm x}_1, {\rm x}_2,{\rm x}_3$, it takes the form
\begin{equation}\label{GG}
 \delta_3 \langle \underline{p}|\mathcal{H}_R|\underline{k}\rangle_{reg}=
\int\frac{dQ \,dQ' } {4\pi^2}
D_0(p,k, Q';R)\,
 {\mathcal{Y}}_{reg}(p,k,Q,Q';m,h),
 \end{equation}
 where
 \begin{equation} \label{DPQ0}
 D_0(p,k,Q, Q';R)=\frac{8 \sin[(p-Q)R/2]\sin[(Q-Q')R/2]\sin[(Q' -k)R/2]}{(p-Q)(Q-Q')( Q'-k)}.
\end{equation}
The function ${\mathcal{Y}}_{reg}(p,k,Q,Q';m,h)$ is  regular near the hyperplane $p=k$, and
vanishes on it 
 \begin{equation}\label{symYreg0}
 {\mathcal{Y}}_{reg}(k,k,Q,Q';m,h)=0
\end{equation}
due to the symmetry 
relation
 \begin{equation*}
 {\mathcal{Y}}_{reg}(p,k,Q,Q';m,h)=-{\mathcal{Y}}_{reg}(k,p,Q',Q;m,h).
\end{equation*}
Exploiting  the  equality
\[
\lim_{R\to\infty} D_0(p,k,Q, Q';R)=8\pi^3\,\delta(p-k)\delta(Q-k)\delta(Q'-k),
\]
one can proceed to the limit $R\to\infty$ in equation \eqref{dppreg1}, which, by virtue of \eqref{symYreg0},
 leads the result  \eqref{RL0}.


 \end{document}